\documentclass[aps,pre,twocolumn,amsmath,amssymb,superscriptaddress]{revtex4-2}
\usepackage{graphicx} 
\usepackage{bbm,overpic}
\usepackage[bbgreekl]{mathbbol}
\usepackage{hyperref}
\usepackage{float}
\usepackage{xcolor,tikz}
\usepackage[utf8]{inputenc}

\definecolor{mypurp}{RGB}{169,83,246}
\definecolor{myyello}{RGB}{229,178,69}
\definecolor{mygree}{RGB}{130,208,67}
\definecolor{myred}{RGB}{202,91,46}
\usetikzlibrary {arrows.meta}
\DeclareRobustCommand\upcirc{%
    \begin{tikzpicture}[baseline=-0.5ex]
        \fill[mypurp] (0,0) circle (0.12cm);
        \draw[-{Classical TikZ Rightarrow[length=.8mm]}, thick, black] (0,0) -- (90:0.22cm);
    \end{tikzpicture}%
}
\DeclareRobustCommand\downcirc{%
    \begin{tikzpicture}[baseline=-0.5ex]
        \fill[myyello] (0,0) circle (0.12cm);
        \draw[-{Classical TikZ Rightarrow[length=.8mm]}, thick, black] (0,0) -- (270:0.22cm);
    \end{tikzpicture}%
}
\DeclareRobustCommand\rightcirc{%
    \begin{tikzpicture}[baseline=-0.5ex]
        \fill[mygree] (0,0) circle (0.12cm);
        \draw[-{Classical TikZ Rightarrow[length=.8mm]}, thick, black] (0,0) -- (0:0.22cm);
    \end{tikzpicture}%
}
\DeclareRobustCommand\leftcirc{%
    \begin{tikzpicture}[baseline=-0.5ex]
        \fill[myred] (0,0) circle (0.12cm);
        \draw[-{Classical TikZ Rightarrow[length=.8mm]}, thick, black] (0,0) -- (180:0.22cm);
    \end{tikzpicture}%
}

\newcommand{\inria}{Universit\'{e} C\^{o}te d'Azur, Inria, CNRS, Calisto team, Sophia Antipolis, France}
\newcommand{\inphyni}{Universit\'{e} C\^{o}te d'Azur, CNRS, Institut de Physique de Nice, France}

\begin{document}
\title{Effects of collective patterns, confinement, and fluid flow on active particle transport}
\author{Chiara Calascibetta}
\affiliation{Department of Physics \& INFN, University of Rome ``Tor Vergata'', Rome, Italy}
\author{Laetitia Giraldi}
\affiliation{\inria}
\author{Zakarya El Khiyati}
\affiliation{\inria}
\author{Jérémie Bec}
\affiliation{\inria}
\affiliation{\inphyni}

\begin{abstract}
The self-organization of active particles on a two-dimensional single-occupancy lattice is investigated, with an emphasis on the effects of boundary confinement and the influence of an external mean fluid flow. The study examines collective behaviors, particularly the transition from a disordered phase to the formation of orientationally ordered patterns, and their impact on particle transport and flux. In the absence of fluid flow, confinement causes particles to accumulate near the walls, leading to clogs or obstructions that hinder movement, or to the formation of bands aligned with the channel. Although these bands limit the particles ability to freely self-propel, they still result in a net flux along the channel. The introduction of an external Poiseuille fluid flow induces vorticity, shifts the phase transition to higher alignment sensitivities, and promotes particle clustering at the channel center, significantly enhancing overall flux.
\end{abstract}

\maketitle


\section{Introduction}
\label{sec:intro}

Ensembles of self-propelled particles provide captivating examples of complex systems that exhibit collective behaviors. Through local interactions, these particles can give rise to remarkable global patterns of movement or organization, such as swarms and flocks. Studying these systems offers insights into the fundamental principles that govern coordinated group behavior across a wide range of contexts, from the collective movements of entities such as active colloids~\cite{mallory2018active}, bacteria~\cite{beer2019statistical}, insects~\cite{kelley2013emergent}, or artificial microswimmers~\cite{yang2021survey} to the synchronized flight of birds~\cite{cavagna2014bird} and fish schooling~\cite{hemelrijk2012schools}. 

The distinct flocking phases that emerge in the self-organization of self-propelled particles share many similarities with the equilibrium phases of matter. Transitions between these phases result from a delicate interplay between energy, entropy, and particle interactions, leading to symmetry breaking within the system.  However, due to their autonomous movements, active particles continuously input energy at a local level, making their collective behavior a celebrated example of self-organization in out-of-equilibrium systems~\cite{toner2005hydrodynamics}. Collective phenomena such as pattern formation, noise-induced phase transitions, and dynamic instabilities are characterized by persisting phase coexistence, long-range correlations, and anomalous fluctuations, often arising from the self-advection of orientational order by individual particles~\cite{chate2020dry}. Understanding and controlling these phenomena have significant implications across various fields, including materials science, biology, and robotics, where harnessing self-organization can lead to the design of novel materials, to efficient transport systems for targeted drug delivery or environmental sensing, or to adaptive strategies for the coordinated movement of swarms of micro-robots.

While much of the work on the collective dynamics of self-propelled particles has focused on open, unbounded domains, many natural and artificial environments involve constrained spaces, such as channels or narrow corridors. Confined geometries can give rise to novel properties not observed in unbounded settings. For instance, confinement can profoundly influence the formation and stability of flocking patterns due to symmetry-breaking effects induced by the geometry of the domain~\cite{negi2023geometry,canavello2024polar}. Moreover, interactions with boundaries can lead to distinctive collective behaviors and spatial structures, such as specific structural organizations~\cite{ribeiro2018lane}, wall accumulations leading to the formation of bottlenecks~\cite{costanzo2012transport}, and even clogging~\cite{caprini2020activity}. These structures play crucial roles in shaping emergent dynamics, particularly when examining how biological systems navigate through complex environments, respond to external stimuli, or work collectively to accomplish specific tasks. Such insights can also inspire new strategies for controlling swarms of artificial microswimmers~\cite{baldovin2023control}.

Understanding how collective patterns influence the transport properties of self-propelled particles is central to many applications. The emergence of swarming or banding order often leads to directional motion and spontaneous particle traffic-like flows, which can significantly enhance transport properties~\cite{bechinger2016active}. Enhanced diffusion of self-propelled particles can further contribute to this effect. However, collective behaviors can also impede movement; for example, clustering and particle interactions may result in phenomena such as absolute negative mobility~\cite{rizkallah2023absolute}. Additionally, geometrical disorder introduced by quenched heterogeneities in the surrounding medium can cause trapping~\cite{chepizhko2013diffusion}, jamming~\cite{reichhardt2014active}, or hinder the transition to collective order in polar flocks~\cite{morin2017distortion}. The influence of external fluid flow on these systems is similarly complex and non-trivial~\cite{caprini2020diffusion}.  Many questions remain open regarding how factors like activity, confinement, and disorder shape the emergent dynamics of active particles, particularly in relation to their global mobility and transport properties. These considerations align with our overarching objective of unraveling the adaptive strategies employed by biological systems to navigate and thrive within complex environments.

The aim of this work is to study the combined effects of confined geometry and external fluid flow on the self-organization and transport properties of self-propelled particles. Among the various numerical models for active matter reviewed in~\cite{shaebani2020computational}, we focus on the restricted four-state active Potts model considered in~\cite{peruani2011traffic}, where active particles are arranged on a two-dimensional square lattice and have four possible states corresponding to four directions of motion. Unlike the unrestricted active Ising or Potts models, which allow for multiple site occupancies~\cite{solon2015flocking,chatterjee2020flocking,agranov2024thermodynamically}, our model restricts each grid point to accommodate only one particle, simulating steric interactions between particles. This volume exclusion is known to facilitate motility-induced phase separation and jamming~\cite{bertrand2022diversity,karmakar2023jamming}. We consider a channel-like domain where particles experience reflective collisions with two horizontal walls and are thus confined in the vertical direction. Additionally, particles are transported and tumbled by a prescribed, external Poiseuille flow~\cite{peng2020upstream}. We explore how boundary conditions and channel size influence the phase transition towards orientational order and investigate the role of fluid flow in shaping collective dynamics and transport properties.

The paper is organized as follows. Section~\ref{sec:model} introduces the model, detailing its parameters and the observables used in our analysis. Section~\ref{sec:noflow} examines the formation of collective patterns in the absence of an external fluid flow, with a focus on how channel geometry affects the various phases that emerge when varying particle density and the strength of inter-particle alignment interactions. Section~\ref{sec:poiseuille} investigates the effects of an imposed Poiseuille flow, analyzing how this influences flocking behavior and its implications for transport properties. Finally, Section~\ref{sec:conclusions} presents concluding remarks and discusses potential future research directions.


\section{Model and observables}
\label{sec:model}

\subsection{A lattice-Vicsek approach}
\label{subsec:lattice}
\begin{figure}[h!]
   \centering
   \includegraphics[width=\columnwidth]{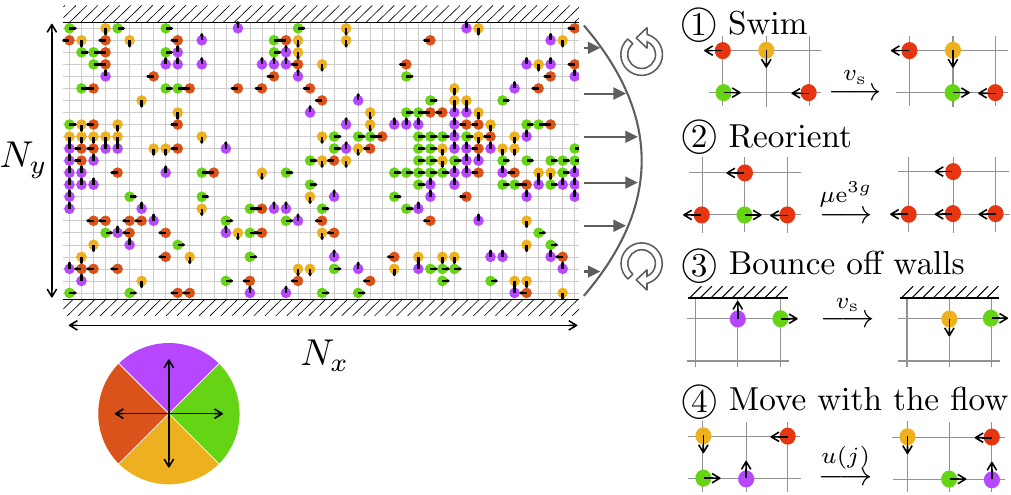}
   \caption{\label{fig:sketch_lattice_channel}\textit{Illustration of the simulation model:} Self-propelled particles on a single-occupancy lattice $N_x\times N_y$, confined between two horizontal walls. Particles can orient in four directions: up \upcirc $\ = (0,1)$, down \downcirc $\ = (0,-1)$, left \leftcirc $\ = (-1,0)$, right \rightcirc $\ = (1,0)$. They can undergo four possible events, as illustrated on the right:
   \raisebox{.5pt}{\textcircled{\raisebox{-.9pt} {1}}}~Swim by hopping to an adjacent unoccupied site at rate $v_\mathrm{S}$;
   \raisebox{.5pt}{\textcircled{\raisebox{-.9pt} {2}}}~Reorient due to angular diffusion, neighbor alignment, or flow rotation;
   \raisebox{.5pt}{\textcircled{\raisebox{-.9pt} {3}}}~Bounce off walls and change orientation;
   \raisebox{.5pt}{\textcircled{\raisebox{-.9pt} {4}}}~Move with the flow at rate $u(j)$, depending on the ordinate $j$.}
\end{figure}

We consider self-propelled particles on a two-dimensional square lattice, where each site can hold at most one particle. A particle indexed by $k\in[1\,\dots\,N_{\rm p}]$ has an orientation $\boldsymbol{p}_{k}$, which can take one of four discrete directions: up, down, left, and right. Particles move to adjacent lattice sites in the direction of their orientation with a rate $v_\mathrm{S}$, provided the target site is unoccupied (event \raisebox{.5pt}{\textcircled{\raisebox{-.9pt} {1}}} depicted in Fig.~\ref{fig:sketch_lattice_channel}). Additionally, particles may reorient to a new direction $\boldsymbol{p}^\prime_{k}$ at a rate $\mu\exp \left[g\,\sum_{\ell\sim k} \boldsymbol{p}^\prime_k \cdot \boldsymbol{p}_{\ell}\right]$, where the summation is over all particles labeled by $\ell$ on neighboring sites of $k$ (event  \raisebox{.5pt}{\textcircled{\raisebox{-.9pt} {2}}}). In the absence of neighboring particles, reorientation occurs randomly at a rate $\mu$. The parameter $g$ quantifies the strength of inter-particle alignment interactions. This setup is consistent with the model studied in~\cite{peruani2011traffic}, originally examined on square periodic domains.

We adopt a channel geometry, confining particles between two horizontal walls at $y=-h$ and $y=h$ (see Fig.~\ref{fig:sketch_lattice_channel}), while applying periodic boundary conditions in the $x$ direction. Particles experience reflective boundary conditions at the walls: if a particle oriented toward a wall attempt to move there at rate $v_\mathrm{S}$, it is reflected back to its original position with a reversed orientation (event \raisebox{.5pt}{\textcircled{\raisebox{-.9pt} {3}}} in Fig.\ref{fig:sketch_lattice_channel}). Similar dynamics of interacting active particles in confined geometries have been explored in both continuous settings~\cite{armbruster2017swarming} and using a hybrid lattice method~\cite{kuhn2021lattice}.

To incorporate the effect of fluid flow, we introduce a systematic translational motion of particles towards $x>0$. This motion is implemented through additional migration events to the right (\raisebox{.5pt}{\textcircled{\raisebox{-.9pt} {4}}} in Fig.~\ref{fig:sketch_lattice_channel}), at a rate $u(j) = U\,[1-(y_j/h)^2]$, dependent on the particle vertical position $y_j =  h(2j/(N_y-1)-1) \in (-h,h)$ and thus their distance from the boundary. This velocity profile mimics Poiseuille flow, describing the creeping motion of a viscous fluid in the channel, with no-slip boundary conditions at the walls. The gradients of the fluid velocity also induce reorientation through an additional rate $\Omega_j = (1/2) |\mathrm{d}u/\mathrm{d}y|(j) = U |y_j|/h^2$, causing particles to rotate either clockwise or counterclockwise, depending on whether they are located in the lower or upper half of the channel.

\begin{table}[h]
    \caption{\label{tab:parameters}Model and simulation parameters.}
    \begin{ruledtabular}
        \begin{tabular}{p{1.6cm}cp{4.2cm}}
            Resolution & $N_y$ & fixed at $25$ \\ \hline
            Alignment sensitivity & $g$ & weight of interactions relative to noise; varied from $0$ to $3.5$\\ \hline
            Migration rate & $v_\mathrm{S}$ & fixed to $100$, faster than noise but comparable to alignment for $g=\mathcal{O}(1)$\\ \hline
            Density & $\rho = N_{\rm p} / (N_x\,N_y)$ & varied in Sec.~\ref{subsec:phases} and then fixed to $0.3$, as a reasonable compromise between ``too dilute'' and ``too packed''\\ \hline
            Box shape & $\lambda = N_x/N_y$ & varied in Sec.~\ref{subsec:phases} from $2$ to $8$, otherwise fixed to $\lambda=2$.\\ \hline
            Flow rate & $U$ & varied in Sec.~\ref{sec:poiseuille}. It is a critical parameter for particle fluxes in the channel.
        \end{tabular}
    \end{ruledtabular}
\end{table}
We normalize the units of length such that $h=1$, and set time units such that the random reorientation rate is $\mu = 1$. Table~\ref{tab:parameters} summarizes the model and simulation parameters after this normalization.

It is important to highlight that the discretized model we use differs significantly from traditional active Ising models. In our approach, we consider volume exclusion effects, meaning that each grid site can be occupied by only one particle, unlike models where multiple particles can coexist at a single location. This single-occupancy constraint effectively represents the finite size of the  particles and has a significant influence on their interactions and self-organization. This rule introduces a natural length scale into the system, which captures both steric and alignment interactions. Furthermore, our model is particularly motivated by practical applications where the channel width is finite, as opposed to the idealized scenario of infinite width ($N_y \to \infty$). In real-world situations, such as microfluidics and biological systems, channels are often finite relative to the size of the particles they contain, and confinement plays a crucial role in the observed collective behaviors. Therefore, our analysis is focused on finite values of $N_y$ to provide insights into the patterns formed by self-propelled particles within such confined domains.

We conduct numerical simulations of this system using a method that integrates particle dynamics on a grid by identifying the next event based on the current state of the particles, rather than advancing the entire system with a fixed time step. This approach, which defines a continuous-time Markov process, allows each particle to undergo events chosen from the first three possibilities mentioned earlier (see also Fig.~\ref{fig:sketch_lattice_channel}) , namely swimming, reorientation, or boundary interactions, according to their associated rates, with an additional rate for reorientation due to fluid vorticity. Advection by the fluid flow is treated separately and is applied to entire segments of particles located at a given $y_j$ ordinate. The next event is drawn from a random selection process based on cumulative rates, where events with higher rates are more likely to be selected.  This strategy aligns with the Doob-Gillespie algorithm, a method widely used in the broader class of dynamic Monte Carlo algorithms for non-equilibrium systems (see, \textit{e.g.}, \cite{gillespie2007stochastic}). By focusing exclusively on the next event, our method optimizes computational efficiency, updating the system only when necessary. However, this approach requires careful consideration of stochastic nature of time progression when performing time averages.

\subsection{Definitions of key observables}
\label{subsec:param}
Our goal is to investigate how confinement and mean flow transport influence the self-organization of the active particle system described above. To characterize different phases, we use the classical \textit{polar order parameter}, defined by the instantaneous mean particle orientation:
\begin{equation}
\Pi = |\boldsymbol{\Pi}|, \quad\mbox{with } \boldsymbol{\Pi}(t) = \frac{1}{N_{\rm p}}\sum_{k=1}^{N_{\rm p}} \boldsymbol{p}_{k}(t).
\label{eq:defPhia}
\end{equation}
The parameter $\Pi$ approaches 0 when particles are randomly oriented or form opposing structures that cancel each other out,  and it approaches 1 when all particles align in the same direction, forming coherent swarms. In the anisotropic channel geometry, it is also insightful to monitor the orientation of the mean vector $\boldsymbol{\Pi}$ to capture directional biases.

Another important quantity for characterizing particle distribution is the \textit{clustering order parameter}:
\begin{equation}
\Psi_4(t) = \frac{1}{4N_{\rm p}}\sum_{k=1}^{N_{\rm p}} \# {\rm Neigh}(k).
\end{equation}
which is close to 0 when particles are dilute and isolated, and close to 1 when they form dense clusters (where almost all particles have four neighbors).

We are also interested in understanding whether the active particles develop any net motion in the channel, either in the direction of the fluid flow ($x>0$) or against it. This is quantified by the \textit{particle streamwise flux}, that is the fraction of particles moving rightward per unit time:
\begin{equation}
\Phi_T = \frac{1}{TN_\mathrm{p}}\!\left(\#\!\!\left[\!\!\begin{array}{c}\text{\small right motions}\\[-4pt]\text{\small in } [t,t+T]\end{array}\!\!\right] - \#\!\!\left[\!\!\begin{array}{c}\text{\small left motions}\\[-4pt]\text{\small in } [t,t+T]\end{array}\!\!\right]\right)\!.
\end{equation}
The long-term average flux is then $\Phi = \lim_{T\to+\infty} \Phi_T$.

Given the non-homogeneous nature of our system, it is valuable to introduce statistics as a function of the distance from the boundary. A natural quantity for this purpose is the \textit{wall-normal density profile}:
\begin{equation}
\rho_j(t) = \frac{1}{N_{\rm p}\,\Delta y}\sum_{k = 1}^{N_{\rm p}} \delta_{Y_k,j},
\end{equation}
with $\Delta y = 2h/N_y$ and such that $\sum_j \rho_j(t) \Delta y = 1$.
Additionally, we extend the various observables to account for their dependence upon $y$ by defining $\boldsymbol{\Pi}_j$  and $\Phi_j$ through partial averages along a given ordinate $j$ on the lattice.

As explained above, our numerical method involves tracking random events, which raises the issue of how to properly address averaging. Our aim is to perform statistical measurements on all observables. We expect that for all parameter values, the system  will eventually reach a statistically stationary and ergodic state after a sufficiently long time. In this regime, physically meaningful statistics should be derived from time averages. Therefore, we focus on time averages, rather than averaging over events. Hereafter, time averages in the asymptotic statistically stationary regime are denoted by angular brackets $\langle\cdot\rangle$. In the sequel, these averages are typically performed over a time of the order of a few thousands after a statistical steady state has been reached.


\section{Swarm statistics in the absence of an external fluid flow}
\label{sec:noflow}

\subsection{Phases, patterns, and transport properties}
\label{subsec:phases}

\begin{figure}[h]
   \centering
   \includegraphics[width=\columnwidth]{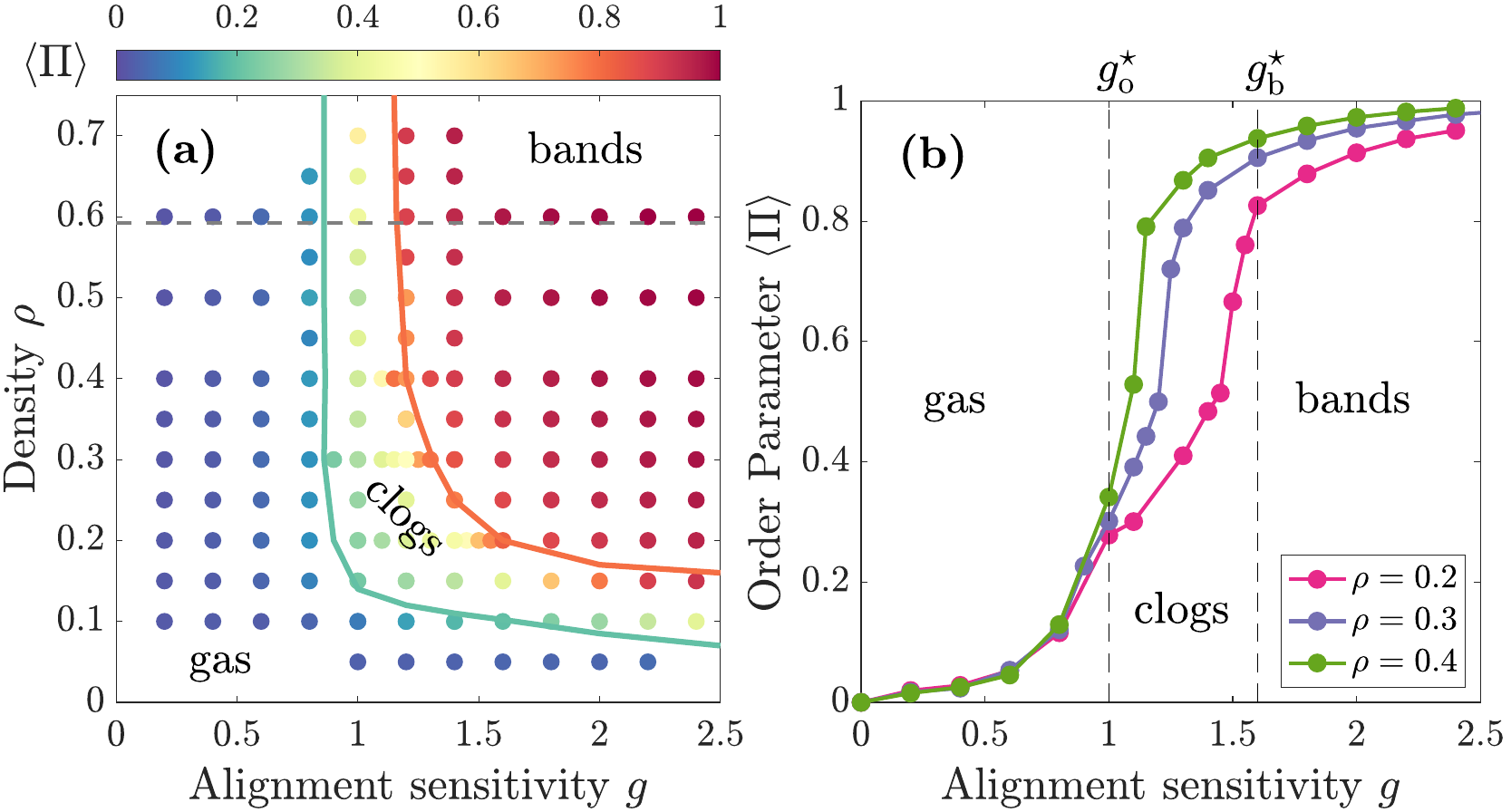}
   \caption{\label{fig:phases_lambda2} \textit{Phase transition without fluid flow}: \textbf{(a)}~Phase diagram in the $(g,\rho)$ parameter space for an aspect ratio $\lambda=2$; colored dots show measured values of the order parameter $\langle\Pi\rangle$, and the solid lines estimate the contour lines for $\langle\Pi\rangle=0.2$  and $0.8$. The horizontal dashed line marks the percolation threshold $\rho\approx 0.592$ on a square lattice. \textbf{(b)}~Average order parameter $\langle\Pi\rangle$ versus $g$ for $\rho = 0.2$, $0.3$, and $0.4$, with critical sensitivities $g_{\rm o}^\star$ and $g_{\rm b}^\star$ also shown. Although $g_{\rm b}^\star$ varies with $\rho$, only its approximate value for $\rho = 0.3$ is plotted for clarity.}
\end{figure}

We begin with characterizing the statistical properties of swarms in the absence of an external fluid flow. To this end, measurements were conducted with $U=0$, while varying the alignment sensitivity $g$ and the total density $\rho$. The results for a channel aspect ratio $\lambda=2$ are presented in Fig.~\ref{fig:phases_lambda2}.  As anticipated, order increases with alignment sensitivity (Fig.~\ref{fig:phases_lambda2}a), transitioning from a gas of point clusters to strongly organized configurations. The dependence upon particle density, shown in Fig.~\ref{fig:phases_lambda2}b, aligns with known results in systems without boundaries~\cite{peruani2011traffic}. In particular, for $g<g_{\rm o}^\star\approx 1$ there is no macroscopic order, independently of the density $\rho$ as long as it is large enough ($\rho\gtrsim0.2$). Polar order occurs at alignment sensitivities above another threshold, \textit{i.e.}\/ for $g>g_{\rm b}^\star$, which displays this time a more visible dependence upon $\rho$. The intermediate phase corresponding to $g_{\rm o}^\star<g<g_{\rm b}^\star$ is characterized by stronger fluctuations of the polar order parameter and to the alternance between different kinds of intermediate structures.

\begin{figure}[h]
   \centering
   \includegraphics[width=.49\columnwidth]{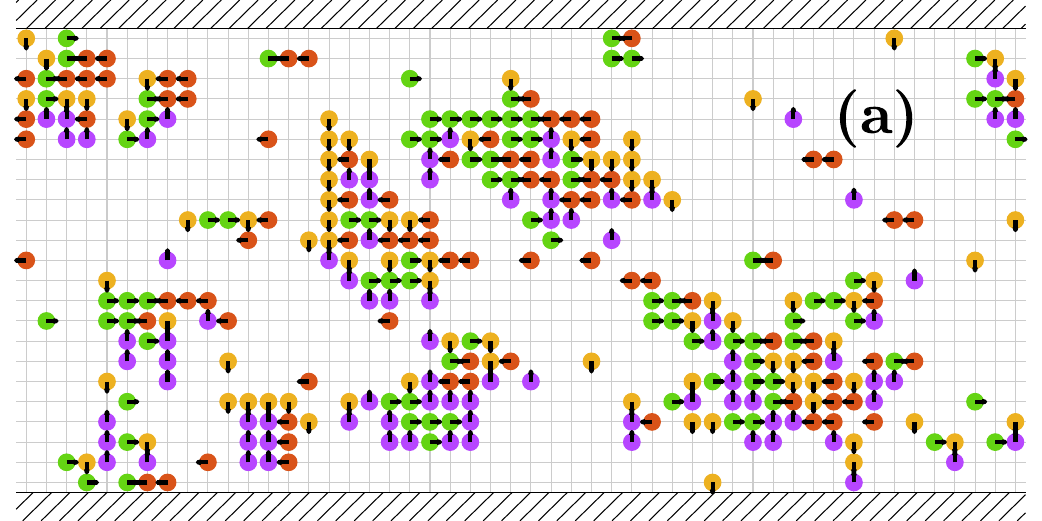} \includegraphics[width=.49\columnwidth]{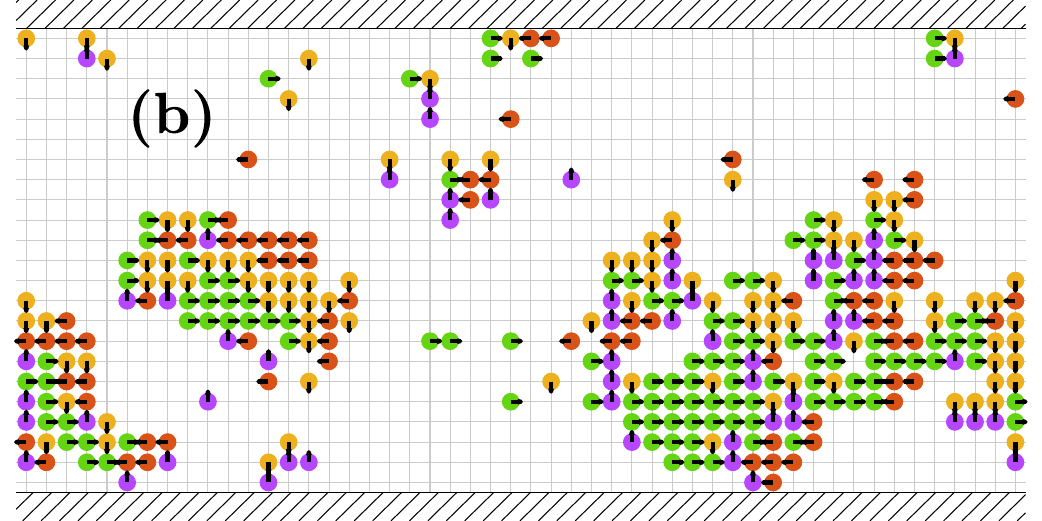}\\[5pt]
   \includegraphics[width=.49\columnwidth]{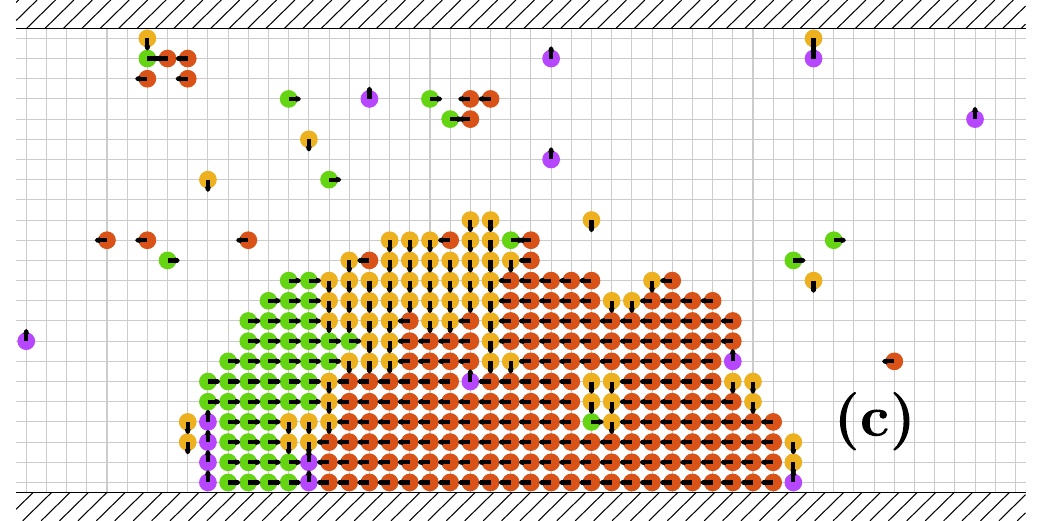} \includegraphics[width=.49\columnwidth]{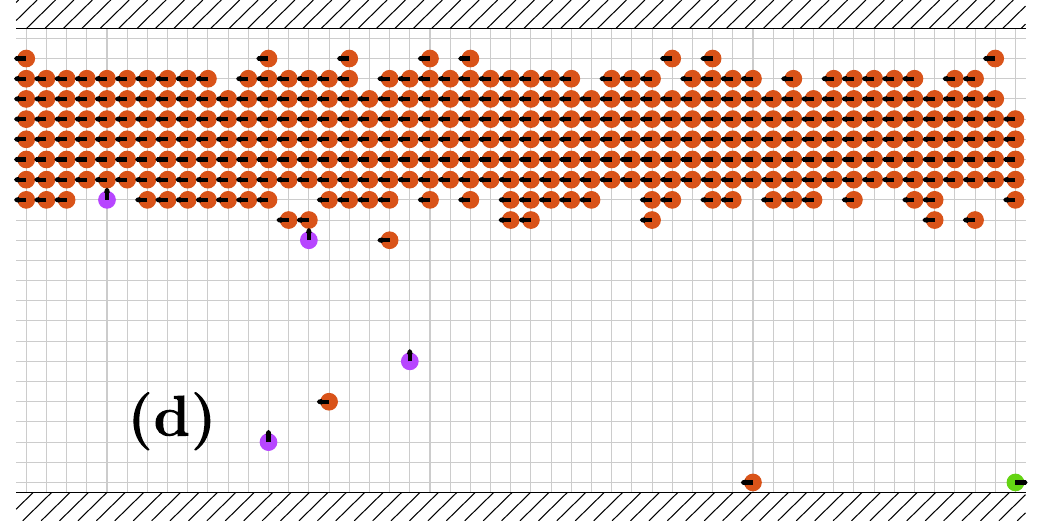}
   \caption{\label{fig:snaps_phases_lambda2} \textit{Illustration of the various phases observed in the absence of fluid flow (for $\rho=0.3$):} \textbf{(a)} and \textbf{(b)} show gas phases (for $g = 0.2$ and $g=0.6$, respectively),  \textbf{(c)} a clog (for $g=1.1$) and \textbf{(d)} a band (for $g=2.8$). }
\end{figure}
Figure~\ref{fig:snaps_phases_lambda2} illustrates the various system configurations that emerge as $g$ is varied. Even at the smallest values of alignment sensitivity (\ref{fig:snaps_phases_lambda2}a and \ref{fig:snaps_phases_lambda2}b), particles form small aggregates whose sizes increase with $g$. The transition to an ordered phase seems thus closely associated with this increase in cluster size, eventually resulting in configurations that span the entire system. This behavior will be further quantified in the subsequent cluster analysis. Qualitatively, the spatial patterns observed in our confined geometry resemble those seen in the purely periodic settings studied in~\cite{peruani2011traffic}, with a key difference: the ``traffic jams'' and ``gliders'' that occur in the intermediate phase $g_{\rm o}^\star<g<g_{\rm b}^\star$ in that study appear only as transients in our case. In our setup, these structures quickly interact with the boundaries, change their nature, and evolve into new types of steady-state configurations that attach to the walls. For instance, figure~\ref{fig:snaps_phases_lambda2}c depicts a structure referred to as a \textit{clog}, which forms and persists for extended periods of time. Such a cluster resembles the traffic jams described in~\cite{peruani2011traffic}, but with the difference that one of the walls contributes to particle blockage, rendering the configuration nearly static while traffic jams diffuse. At higher values of alignment sensitivity ($g>g_{\rm b}^\star$), the self-propelled particles organize into bands where they all align in the same direction. The channel topology imposes geometric constraints on such clusters, causing particles to align parallel to the walls and form band along the periodic direction, as illustrated in Fig.~\ref{fig:snaps_phases_lambda2}d. While these various observations suggest parallels with periodic settings, they also highlight the critical role of boundaries in determining the details of the system's ultimate phase behavior.

\

\begin{figure}[t]
   \centering
   \includegraphics[width=\columnwidth]{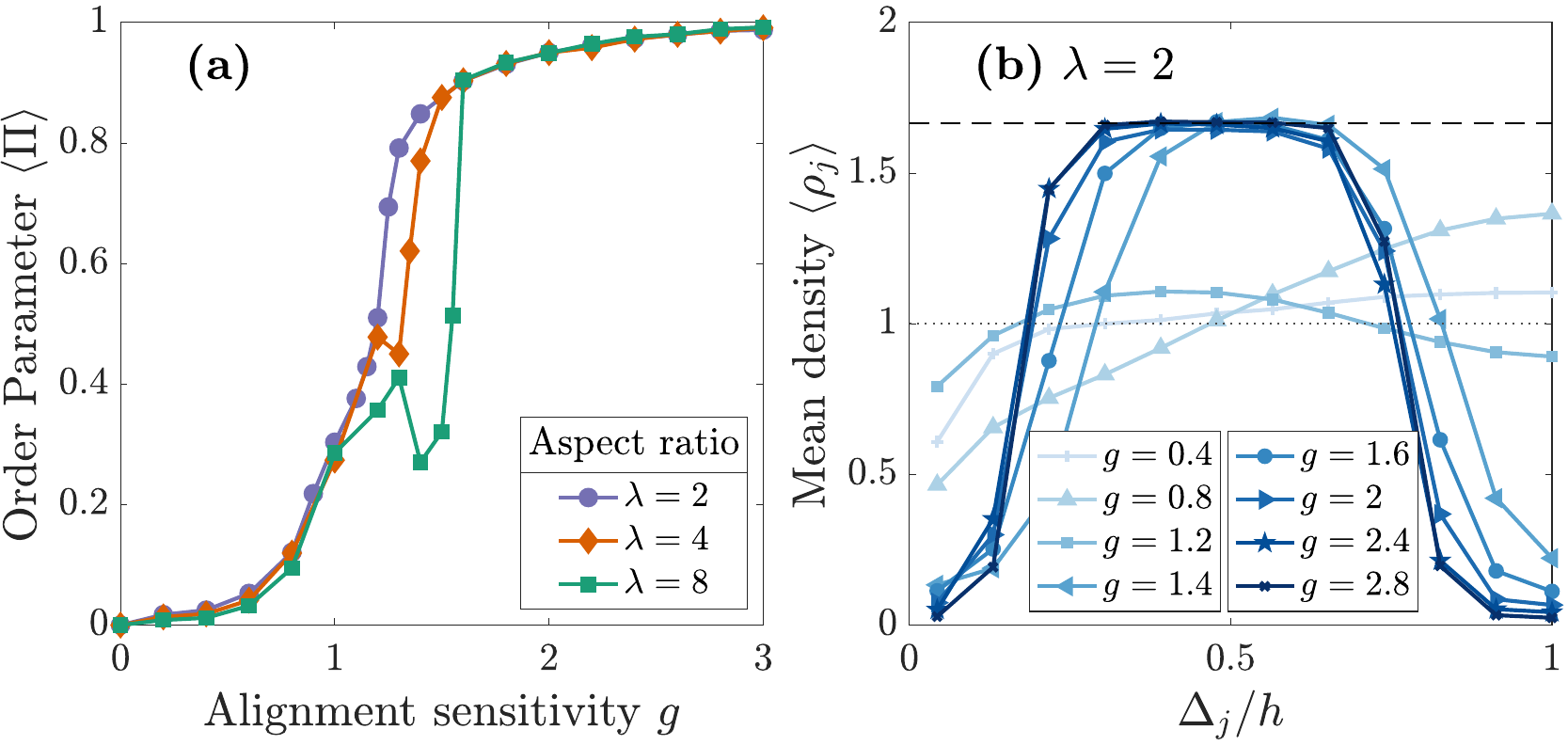}
   \caption{\label{fig:phases_var_lambda_and_density} \textbf{(a)}~Average order parameter $\langle\Pi\rangle$ as a function of $g$ for density $\rho=0.3$ and various aspect ratios $\lambda$ of the simulation domain. \textbf{(b)}~Average density as a function of the distance from the boundary $\Delta_j = \min(y_j+h,h-y_j)$ for $\lambda=2$ and various values of $g$. The density is normalized such that $\langle\rho_j\rangle\equiv 1$ for a homogeneous distribution (dotted line); dense packing then corresponds to $\langle\rho_j\rangle\equiv h/\rho$ (dashed line).}
\end{figure}

We next explore the effects of varying the aspect ratio $\lambda$ of the channel while maintaining a constant resolution in the $y$ direction. Increasing the aspect ratio introduces new phases that persist for extended periods of time before eventually transitioning to a steady state. This phenomenon is accompanied with pronounced fluctuations in the order parameters near the gas-to-band phase transition. Figure~\ref{fig:phases_var_lambda_and_density}a shows the average order parameter $\langle \Pi \rangle$ as a function of $g$ for different aspect ratios $\lambda \in \{ 2, 4, 8 \}$. At sufficiently low and high values of $g$, the curves overlap, indicating that the polar order parameter is there independent of $\lambda$.   Bands form above a critical value of the alignment sensitivity $g>g_{\rm b}^\star(\lambda)$, which significantly increases as a function of the aspect ratio. This is again consistent with the findings of \cite{peruani2011traffic} where the effect of domain size on such a transition was investigated in periodic settings. However, near the transition point, an increase in $\lambda$ leads to greater fluctuations in the order parameter, which can be interpreted as resulting from an increased difficulty for the system to reach a statistical steady state at larger aspect ratios.

Figure~\ref{fig:phases_var_lambda_and_density}b shows the steady-state wall-normal density profiles $\langle\rho_j\rangle$ obtained for a global density $\rho=0.3$, with $\lambda=2$, and various alignment sensitivities $g$. The data are plotted as a function of the distance from the boundaries, as the statistics are symmetric with respect to this distance. At low values of $g$, particles remain dilute and are almost uniformly distributed across the channel. As $g$ increases, the active particles begin to cluster into bands that move closer to the boundaries, eventually located within a few lattice sites from the walls. 

\begin{figure}[h]
	\centering
	\includegraphics[width=\columnwidth]{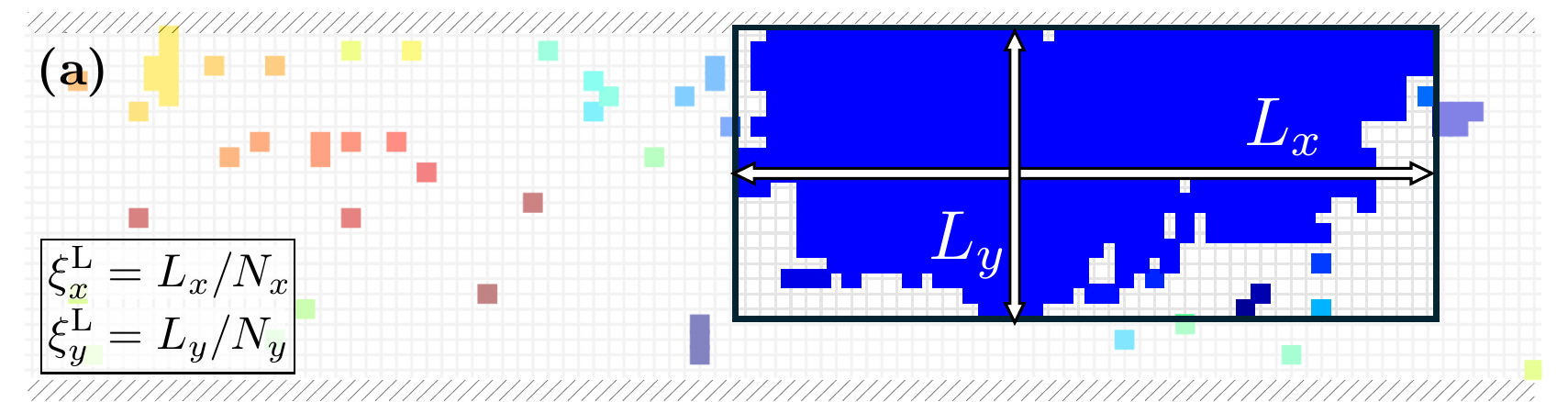}\\[10pt]
	\includegraphics[width=\columnwidth]{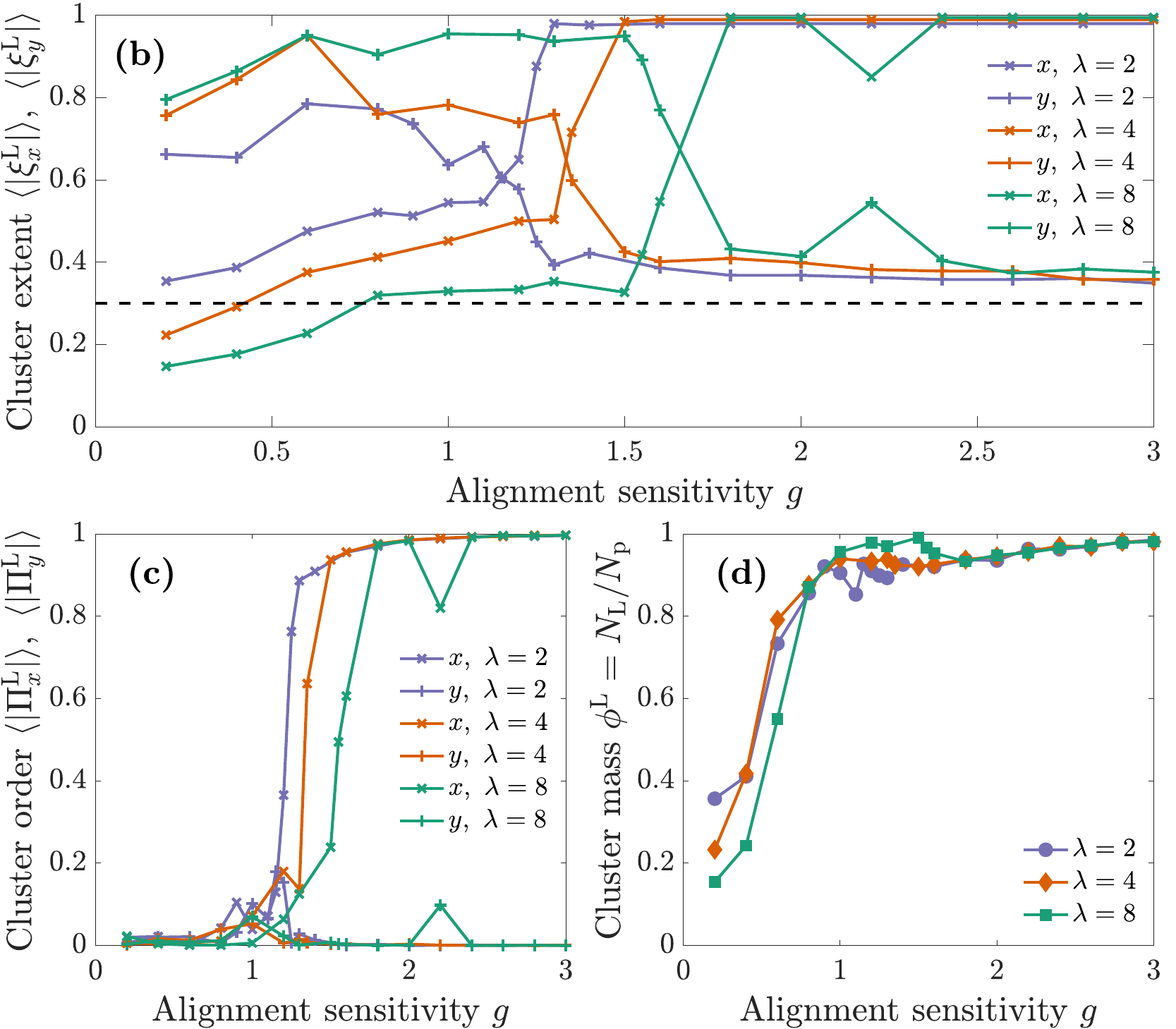}
	\caption{\label{fig:cluster_analysis} \textit{Cluster analysis.}  \textbf{(a)} Different colors represent distinct clusters in the domain, with the largest highlighted in blue. The maximum extents of this cluster, $L_x$ and $L_y$, are then calculated, and $\xi_x^{\text L}$ and $\xi_y^{\text L}$ represent these extents normalized by the domain size.  \textbf{(b)} Mean maximum extents, $\langle \xi_x^{\text L} \rangle$ and $\langle \xi_y^{\text L} \rangle$, as a function of $g$ for various aspect ratios $\lambda$ of the simulation domain. The dashed line corresponds to the expected values for the band configuration: $\langle \xi_y^{\text L}\rangle = \rho = 0.3$.  \textbf{(c)} Mean order parameters $\langle \Pi_x^{\text L} \rangle,\, \langle \Pi_y^{\text L} \rangle$ as function of $g$.  \textbf{(d)} Normalized cluster density $\phi^{\text L}$ as function of $g$. In all plots, full symbols represent statistics for the $x$ direction, while open symbols represent statistics for the $y$ direction.}
\end{figure}

To further understand the convergence to a steady state, we conduct a cluster analysis. For all $g$ and $\lambda$ values, we identify as a function of time the largest cluster within the domain, defined as the cluster encompassing the most particles (see Fig.~\ref{fig:cluster_analysis}a). Table~\ref{tab:cluster} summarizes the variables used in our analysis. As shown in Figs.~\ref{fig:cluster_analysis}b and c, a phase transition is observed in both the maximum cluster extents ($\xi_x^{\text L}$ and $\xi_y^{\text L}$) and the local order parameters ($\Pi_x^{\text L}$ and $\Pi_y^{\text L}$) at higher $g$ values as $\lambda$ increases. At lower $g$ values, the difference between $\xi_x^{\text L}$ and $\xi_y^{\text L}$ becomes more pronounced with increasing $\lambda$. This suggests that smaller aspect ratios favor the formation of clogs, while larger ratios lead to the formation of obstructions, where clogs extend between the two walls of the channel.

\begin{table}[b]
	\caption{\label{tab:cluster}Variables for cluster analysis.}
	\begin{ruledtabular}
		\begin{tabular}{p{3cm}cp{3cm}}
			Normalized \text{cluster~~~} mass & $\phi^{\text L} = N_{\rm L} / N_{\rm p}$ &  Fraction of particles in the biggest cluster. \\ \hline
			Normalized maximal extents & $\xi_x^{\text L}, \xi_y^{\text L}$ & Estimated as shown in Fig.~\ref{fig:cluster_analysis}a. \\ \hline
			Cluster \text{order~~~~~~~~~~} \text{parameters} & $\Pi_x^{\text L}, \Pi_y^{\text L}$ & evaluated within the biggest cluster. 	
		\end{tabular}
	\end{ruledtabular}
\end{table}
For  $g$ values above the transition, the maximum cluster extents converge to $ \xi_x^{\text L} \rightarrow 1$ and $ \xi_y^{\text L} \rightarrow \rho $, regardless of $\lambda$, matching the expected values for a band configuration with a single dominant cluster. This is further quantified in Fig.~\ref{fig:cluster_analysis}d, which shows the density of the largest cluster. In the ordered phase, the cluster size increases and attains $\phi^{\text L} \approx 1 $, indicating that all particles belong to the same cluster. Accordingly, the order parameters transition from a near zero mean at low $g$ values to $\Pi_x^{\text L} \approx 1$ and $\Pi_y^{\text L}\approx 0 $ in the band configuration, as shown in Fig.~\ref{fig:cluster_analysis}c.

\

\begin{figure}[h]
 	\centering
 	\includegraphics[width=\columnwidth]{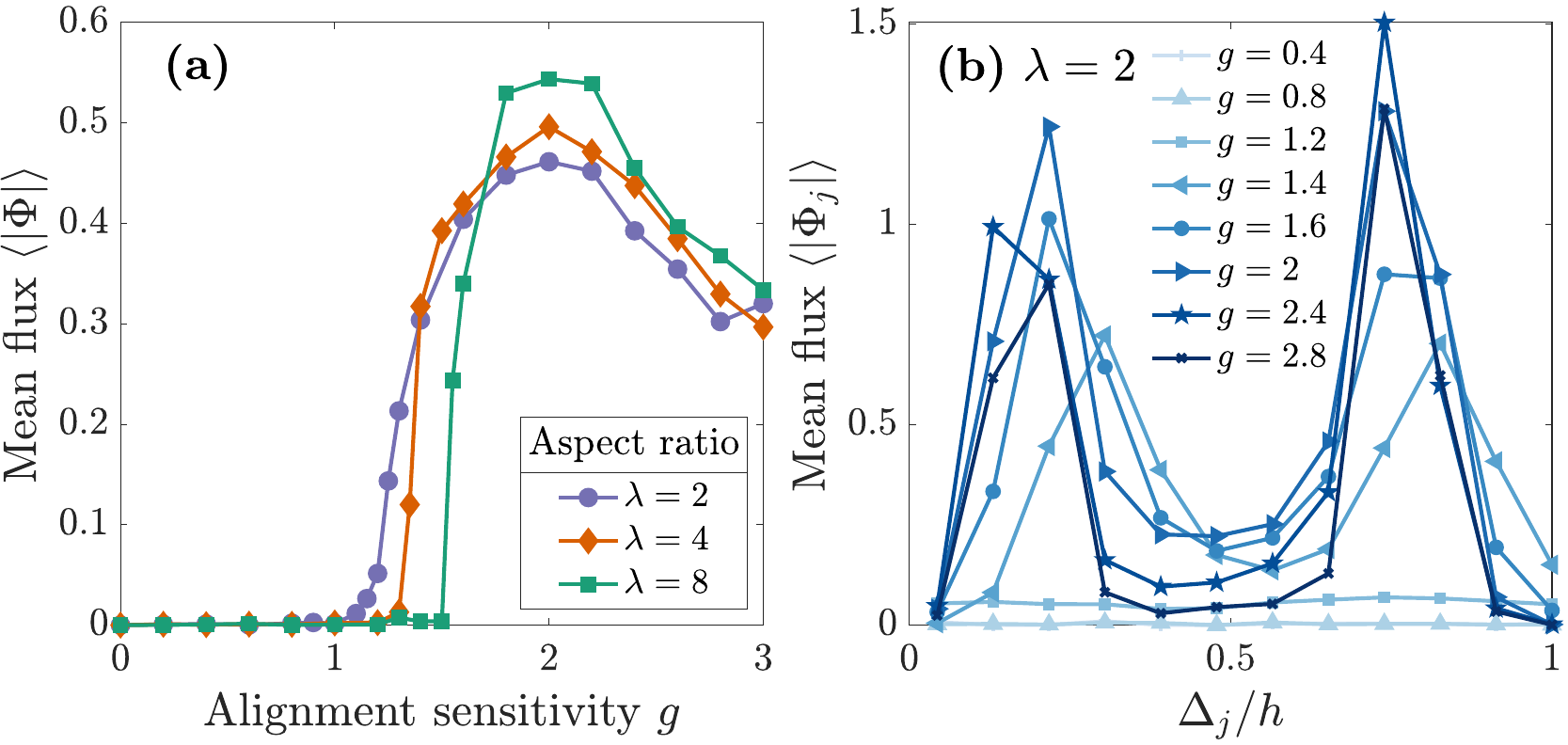}
	\caption{\label{fig:fluxes_var_lambda}  \textbf{(a)}~Average magnitude of the flux $\langle|\Phi|\rangle$ as a function of $g$ for density $\rho=0.3$ and various aspect ratios $\lambda$. \textbf{(b)}~Absolute value of the mean flux as a function of the distance from the boundary $\Delta_j = \min(y_j+h,h-y_j)$ for $\lambda=2$, and various alignment sensitivities.}
 \end{figure}
Finally, we examine the particle fluxes in the different phases described above. Figure~\ref{fig:fluxes_var_lambda}a shows the average magnitude of the streamwise displacement $\Phi$ as a function of $g$. Like polar order, the net flux undergoes a clear phase transition from a near-zero mobility to a finite value. This transition occurs at alignment sensitivities corresponding to the formation of bands, \textit{i.e.}\/ at $g = g_{\rm b}^\star(\lambda)$, as indicated by the comparison with Fig.~\ref{fig:phases_var_lambda_and_density}a. The fact that higher alignment sensitivities are required for increasing aspect ratios $\lambda$ reinforces the idea that the transition to the flux is controlled by the critical sensitivity $g_{\rm b}^\star(\lambda)$ for band formation. In other words, longer channels demand stronger alignment sensitivity to achieve similar ordered states. Consequently, significant particle transport occurs only when the particles self-organize into a band.

The connection between band formation and net flux becomes even more evident when examining the wall-normal profiles of particle fluxes, as shown in Fig.~\ref{fig:fluxes_var_lambda}b for $\lambda=2$. These profiles reveal a competition between two effects. On the one hand, polar order generates a net flux by providing a directional bias to particle motion.  This is particularly apparent in the sharp evolution of the profiles between $g=1.2$ and $g=1.4$. On the other hand, volume exclusion effects and too dense clustering impede the particles' ability to move efficiently at their swimming speed. Hence, as $g$ increases well above the threshold, particles become more densely packed and stacked within the cluster cores, leading to a reduction in flux. However, migration continues at the edges of the band, where the particle density is lower. In these regions, the particles can move parallel to the structure, as observed for instance in Fig.~\ref{fig:snaps_phases_lambda2}d. 

This interplay between alignment and clustering results in the non-monotonic dependence of $\langle|\Phi|\rangle$ on $g$ observed in Fig.~\ref{fig:fluxes_var_lambda}a. Above the transition, the average particle mobility increases rapidly, reaching a maximum around $g\approx 2$, before gradually decreasing at higher alignment sensitivities. Before this peak, increasing $g$ enhances alignment and improves the bias in flux. Beyond this peak, clustering becomes dominant, leading to denser clusters where particles are stacked, and the flux subsequently diminishes. The maximum flux value increases with the aspect ratio $\lambda$, likely because longer channels, and thus longer bands, allow for greater density fluctuations at the edges, leading to more frequent particle migrations. 

\subsection{Transport during transients and transitions}
\label{subsec:transients}

We have seen in the previous subsection that steady-state particle distributions can lead to a net average flux within the channel. This occurs when particles organize into a band, specifically for $g>g_{\rm b}^\star$, with transport primarily taking place at the edges of these patterns. However, at intermediate times and for alignment sensitivities $g$ close to the transition, transport exhibits significant fluctuations, both large and small. As we will now discuss, these fluctuations are caused by the formation and subsequent breakdown of intermediate structures, such as clogs or obstructions.

\begin{figure}[h]
   \centering
   	\includegraphics[width=\columnwidth]{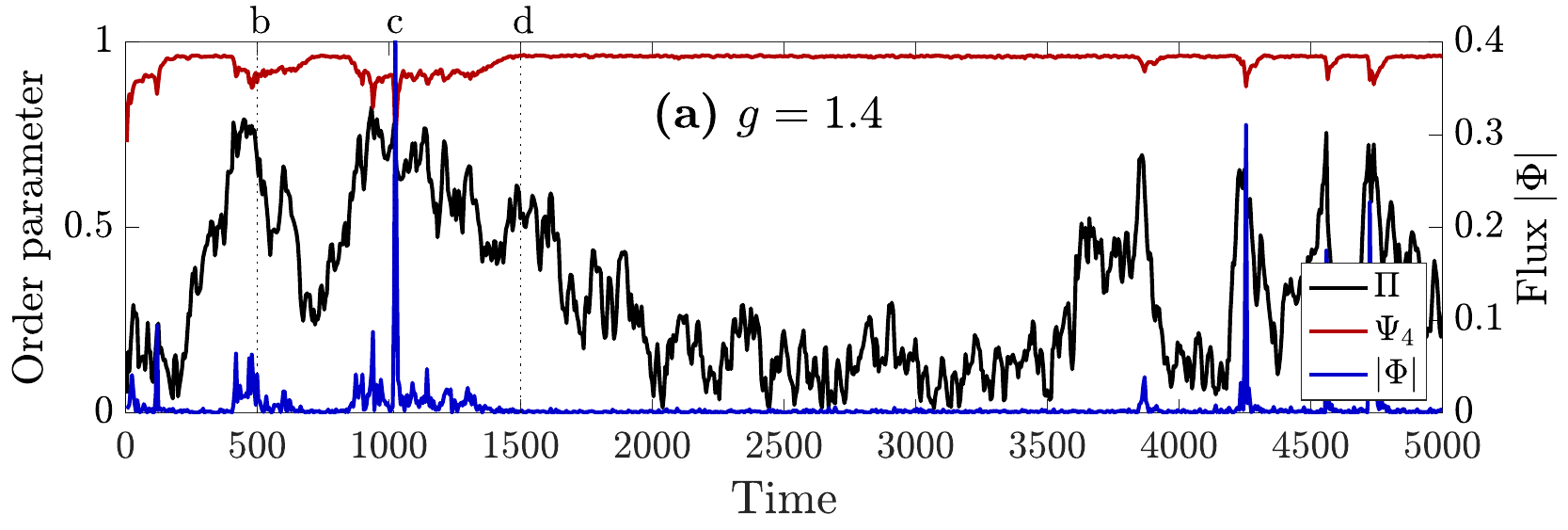}
   	\includegraphics[width=\columnwidth]{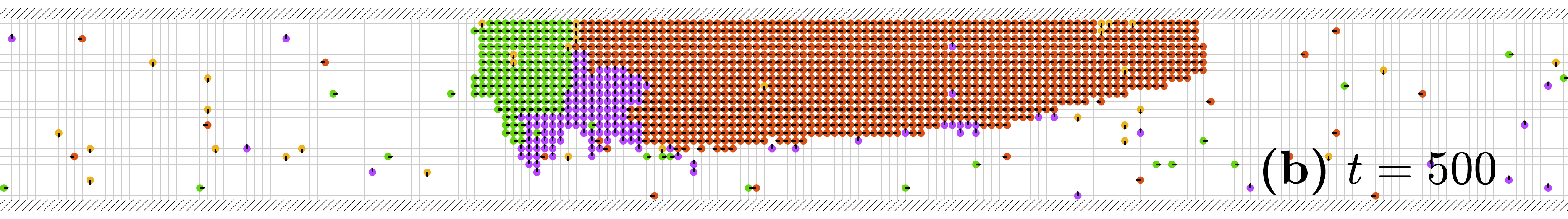}
   	\includegraphics[width=\columnwidth]{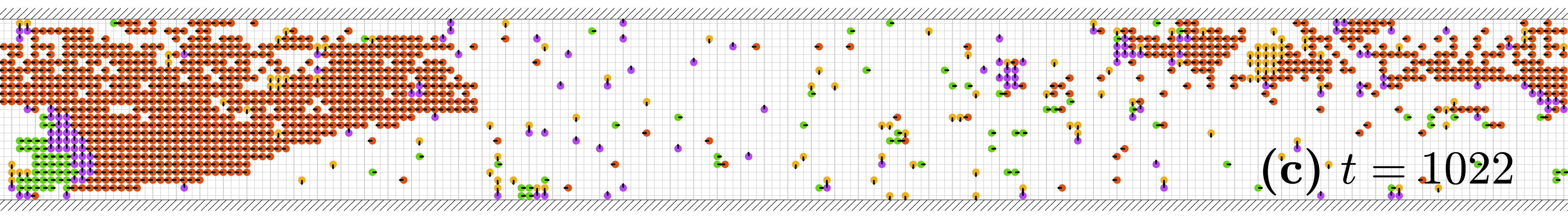}
   	\includegraphics[width=\columnwidth]{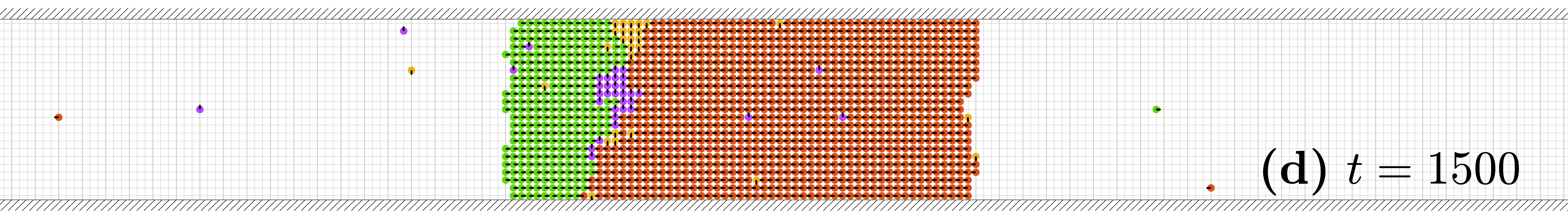}
	\vspace{-15pt}
   	\caption{\label{fig:noconvergence_smaller_g} \textit{Fluctuations below the phase transition}\,---\,here for $g=1.4$ and $\lambda=8$. \textbf{(a)}~Order parameters $\Pi$ and $\Psi_4$ and flux magnitude $|\Phi|$ as a function of time. Representative configurations at times \textbf{(b)}~$t=500$ when particles form a clog, \textbf{(c)}~$t=1022$ at which a large fluctuation of flux is observed, and \textbf{(d)}~$t=1500$ when particles form an obstruction.}
\end{figure}
As previously mentioned, alignment sensitivities $g$ just below the transition, combined with a large aspect ratio $\lambda$, lead to highly fluctuating macroscopic quantities. Figure~\ref{fig:noconvergence_smaller_g}a shows the time evolution of the polar and clustering order parameters, together with the flux amplitude, for $g = 1.4 < g_{\rm b}^\star(\lambda)$ and $\lambda = 8$. Despite running the simulation for much longer durations compared to other $g$ values, a steady state remains elusive. Figures~\ref{fig:noconvergence_smaller_g}b--d illustrate various configurations of the system. At time $t=500$ (Fig.~\ref{fig:noconvergence_smaller_g}b), a clog is present in the particle distribution. It is followed by a surge in flux (Fig.~\ref{fig:noconvergence_smaller_g}c), during which all particles align in the same direction, yet free from interference with neighbors. However, instead of transitioning to a stable band configuration, the particles rather form an obstruction (Fig.~\ref{fig:noconvergence_smaller_g}d). Notably, the new obstruction forms nearly in the same location as the initial clog, though this happens to be coincidental. During the time between the clog's dissolution and the obstruction's formation, most particles have traveled multiple times around the channel. This cycle recurs over time, so that clogs, obstructions, and dilute particles do not constitute stable patterns at alignment sensitivities below the phase transition. The net flux exhibits at these parameter values highly intermittent statistics, as inferred from the time profile of $|\Phi|$ in Fig.~\ref{fig:noconvergence_smaller_g}a. The probability distributions of the instantaneous flux, not shown here, seem to display super-exponential tails. However, accurately measuring them would require extremely long statistics, which is beyond the scope of this study.

\begin{figure}[h]
	\centering
   	\includegraphics[width=\columnwidth]{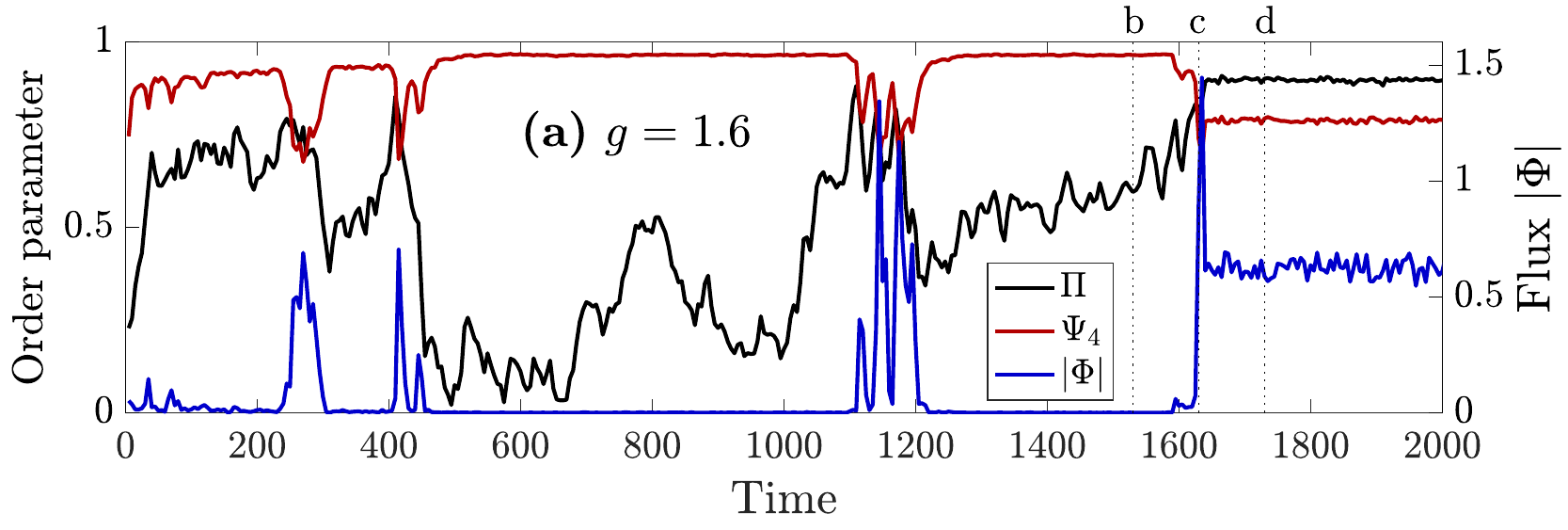}
   	\includegraphics[width=\columnwidth]{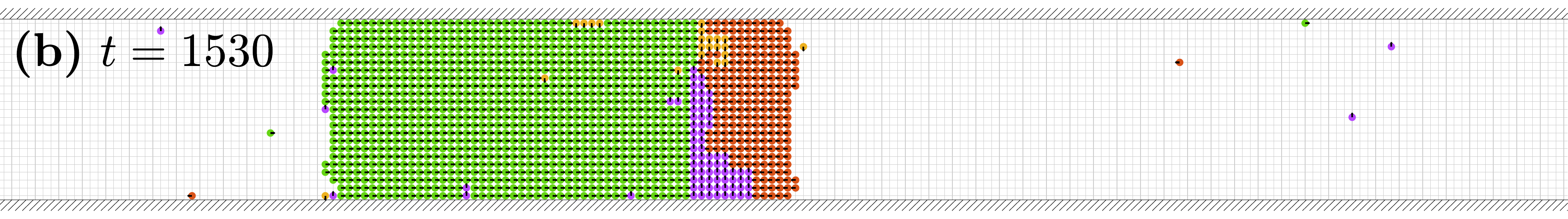}
   	\includegraphics[width=\columnwidth]{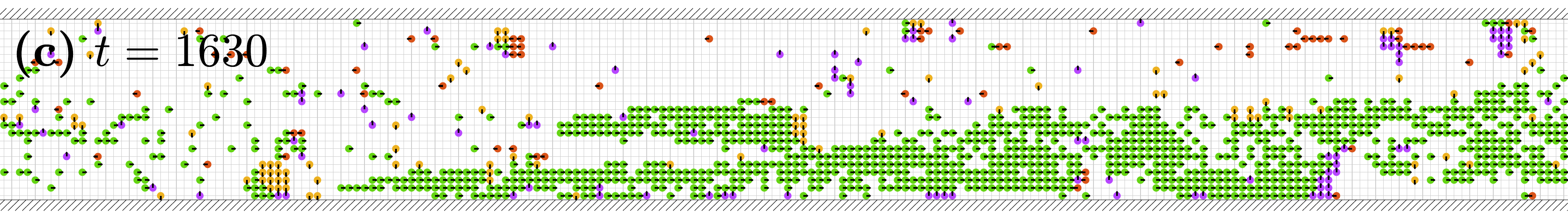}
   	\includegraphics[width=\columnwidth]{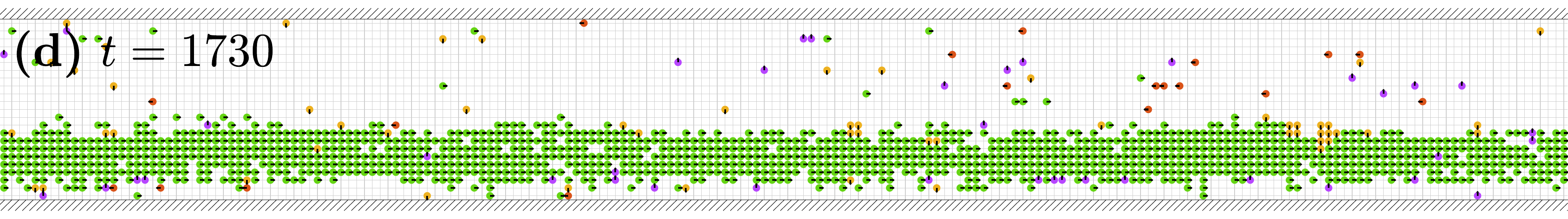}
   	\caption{\label{fig:long_transient} \textit{Transients above the phase transition}\,---\,here for $g=1.6$ and $\lambda=8$. \textbf{(a)}~Time evolution of the polar and clustering order parameters, and of the flux magnitude. Convergence to the band configuration (final plateaus) is preceded by long periods without any flux \textbf{(b)}. The final transition to order goes through a burst of flux \textbf{(c)} where particles are aligned in the same direction without yet being blocked by neighbors in their motion as in the final band configuration \textbf{(d)}}
\end{figure}
A similar behavior is observed for alignment sensitivities above the transition, during the transient regime preceding the formation of a stable band. Figure~\ref{fig:long_transient}a shows again the time evolution of the order parameters $\Pi$ and $\Psi_4$, along with the flux magnitude $|\Phi|$. Before the system reaches the final band configuration (indicated by plateaus in the global quantities at $t\gtrsim1650$), it undergoes prolonged phases of obstruction. These configurations are characterized by $\Psi_4$ values close to 1 (indicating densely packed particles), a complete absence of flux, and a slowly varying polar order parameter $\Pi$. As illustrated in Fig.~\ref{fig:long_transient}b, such transitional stages appear to be associated with the propagation of a front between left- and right-moving particles until one direction predominates. Notably, between $t=1200$ and $1600$, there is an almost linear increase in $\Pi$, while $\Psi_4$ and $|\Phi|$ remain constant. Just before the final transition to the ordered band configuration, the flux exhibits a strong peak. As seen in Fig.~\ref{fig:long_transient}c, many particles align in the same direction \rightcirc, temporarily free from interference by neighbors, allowing for significant migration. Eventually, the particle settle in a band (Fig.~\ref{fig:long_transient}d), and the flux decreases accordingly.

\begin{figure}[h]
   	\centering
   	\includegraphics[width=\columnwidth]{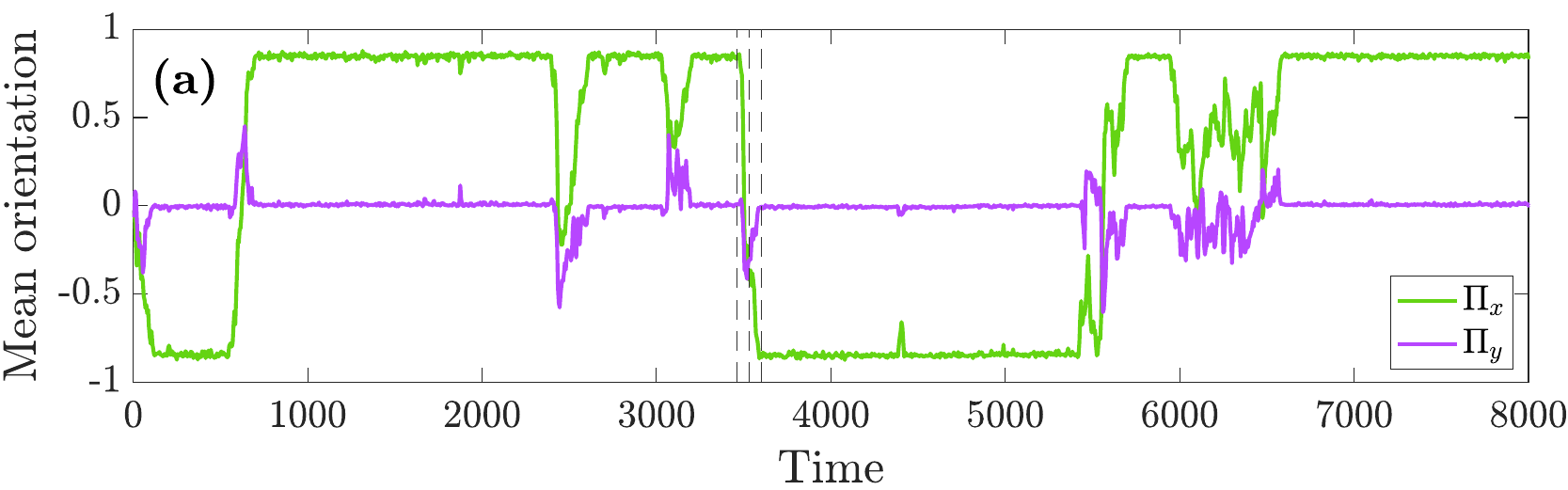}
   	\includegraphics[width=\columnwidth]{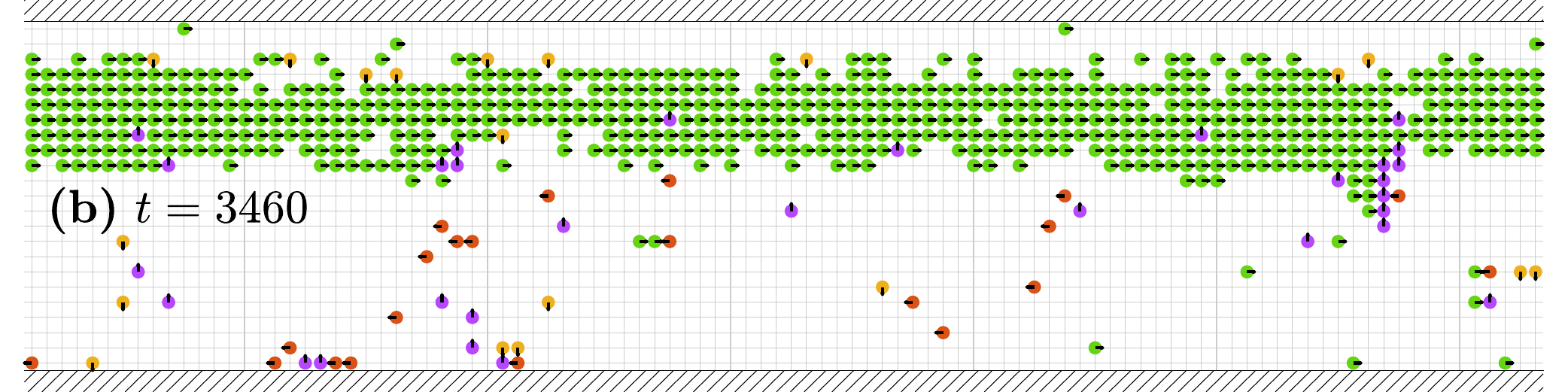}
   	\includegraphics[width=\columnwidth]{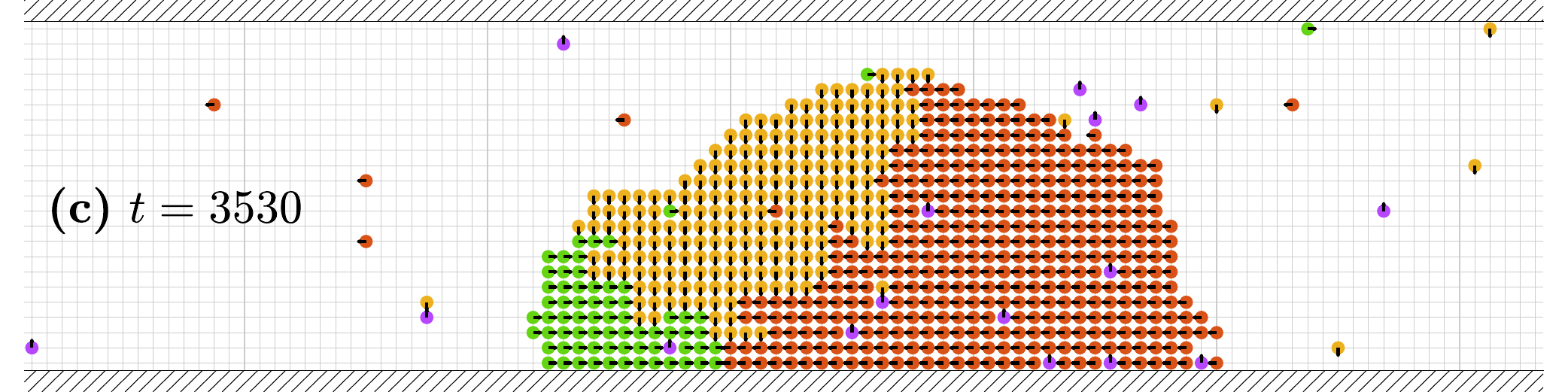}
   	\includegraphics[width=\columnwidth]{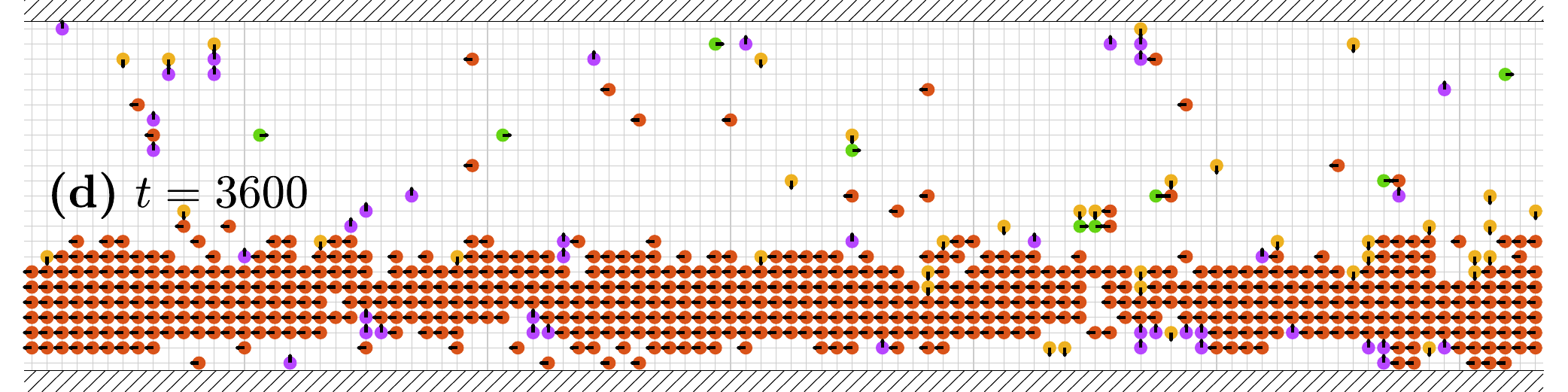}
	\caption{\label{fig:long_time_g_1.4} \textit{Dynamical transition:} \textbf{(a)}~Time evolution of the horizontal and vertical components $\Pi_x$ and $\Pi_y$ of the mean orientation for $\lambda=4$ and $g=1.4>g_{\rm b}^\star(\lambda)$. Around $t=3500$, the particles transition from a right-oriented band \textbf{(b)} to a left-oriented one \textbf{(d)}, forming temporarily a constriction at intermediate times \textbf{(c)}.}
\end{figure}
To complete this picture, it is important to note that in the statistical steady state, the inherent stochasticity of our system can lead to dynamical transitions in its global configuration. These transitions involve for instance all particles changing orientation from left to right or migrate from one half of the channel to the other.  An example of such a transition is shown in Fig.~\ref{fig:long_time_g_1.4}, where particles, initially aligned to the right in a band configuration along the top wall, suddenly migrate to the lower wall and reorient to the left. During this transition, they form a temporary clog along the bottom wall. The time evolution of the horizontal and vertical components of the mean orientation vector $\boldsymbol{\Pi}$ is shown in Fig.~\ref{fig:long_time_g_1.4}a, highlighting the abrupt nature of this shift. It is also evident that the system can remain in a given configuration for very long periods, with transitions between macroscopic states occurring only infrequently, potentially over thousands of time units.


\section{Self-organization and transport in the presence of a Poiseuille flow}
\label{sec:poiseuille}

In this section, we investigate the influence of an external fluid flow on the orientational and transport behavior of the self-propelled particles. Specifically, we examine the effects of a Poiseuille flow, characterized by a parabolic velocity profile, on particle dynamics in the channel geometry. The relative strength of the flow is determined by the dimensionless amplitude parameter $U/v_{\rm S}$, with the fluid acting to transport the particle at speeds dependent upon their distance from the boundary and to rotate them at a rate determined by the local vorticity (as detailed in Sec.~\ref{subsec:lattice}). We present the results from a series of numerical simulations in which both the alignment sensitivity $g$ and the flow parameter $U$ are systematically varied, while other parameters are kept constant, specifically the total particle density at $\rho=0.3$ and the channel aspect ratio at $\lambda=2$.

\begin{figure}[h]
	\centering
	\includegraphics[width=\columnwidth]{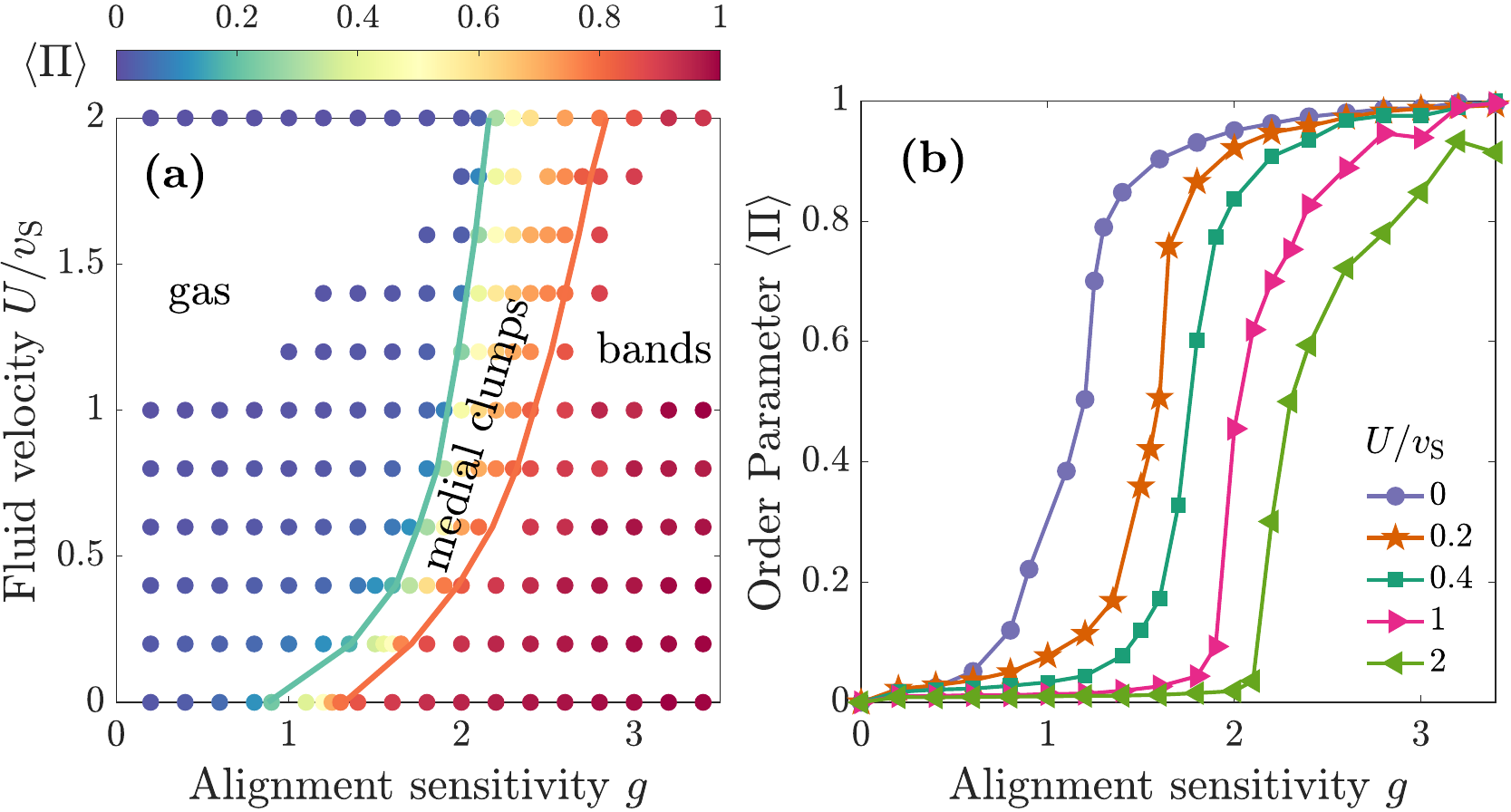}
	\caption{\label{fig:phases_var_v1} \textit{Phase transition in the presence of a Poiseuille fluid flow} (for a density $\rho=0.3$ and an aspect ratio $\lambda=2$). \textbf{(a)}~Phase diagram in the parameter space $(g,U/v_{\rm S})$; colored dots show measured values of the order parameter $\langle\Pi\rangle$, while the two solid lines are estimates of the contour lines $\langle\Pi\rangle =0.2$ and $\langle\Pi\rangle =0.8$. \textbf{(b)}~Average order parameter $\langle\Pi\rangle$ as a function of $g$ for various values of the fluid velocity $U$ as labeled. }
\end{figure}
We expect the presence of a Poiseuille flow to weaken the orientational interactions between particles. Due to the spatially varying vorticity, the fluid flow imposes different rotational rates to particles in neighboring cells along the wall-normal direction $y$. Consequently, they will align in the same direction only if the alignment sensitivity is sufficiently strong to overcome the influence of the flow. This effect can be clearly seen in the phase diagram shown in Fig.~\ref{fig:phases_var_v1}a, where the transition to an orientationally ordered phase occurs at higher values of $g$ as the fluid flow amplitude $U$ increases. The two critical alignment sensitivities\,---\,$g_{\rm o}^\star$, below which there is no macroscopic order, and $g_{\rm b}^\star$, above which particles organize into bands\,---\,are approximately indicated by the contour lines $\langle\Pi\rangle =0.2$ and $\langle\Pi\rangle =0.8$, respectively. Both curves exhibit an exponential dependence of $U/v_{\rm S}$ on $g$, reflecting the balance between the flow-induced tumbling rate $\propto U/h$, and the rate $\exp(n\,g)$ at which $n$ aligned neighbors influence the orientation of a given particle. For the lower critical sensitivity $g_{\rm o}^\star$, numerics suggest that $U \propto \exp(3g_{\rm o}^\star)$, and thus that interactions with three neighbors dominate the apparition of macroscopic order. For the upper critical sensitivity $g_{\rm b}^\star$, one observes that $U \propto \exp(2g_{\rm b}^\star)$, indicating that the formation of bands depends on how two aligned particles influence a third. Consequently, the gap between $g_{\rm o}^\star$ and $g_{\rm b}^\star$ widens logarithmically with increasing $U$. This is evident in Fig.~\ref{fig:phases_var_v1}b, which shows that higher flow velocities lead to a broader intermediate phase.

\begin{figure}[h]
	\centering
	\includegraphics[width=.49\columnwidth]{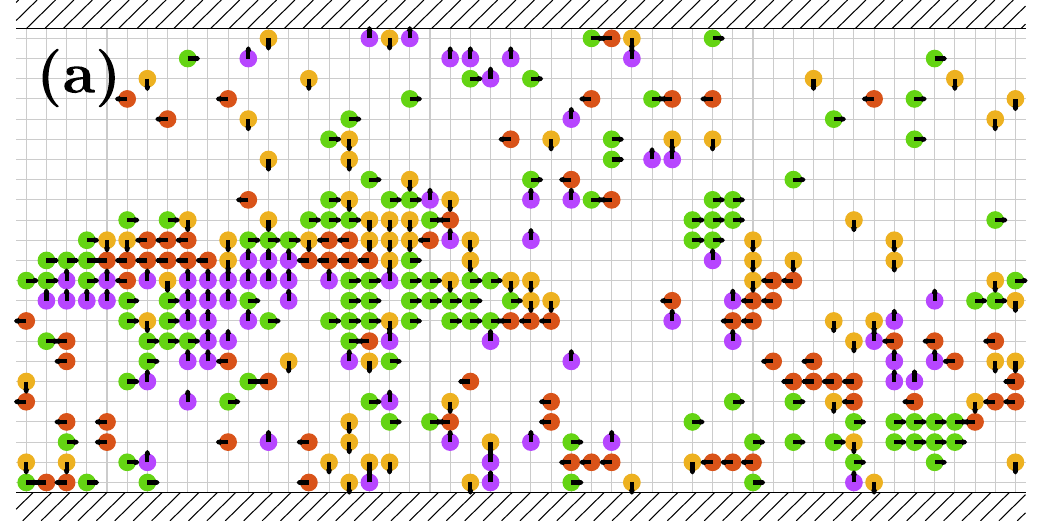}
	\includegraphics[width=.49\columnwidth]{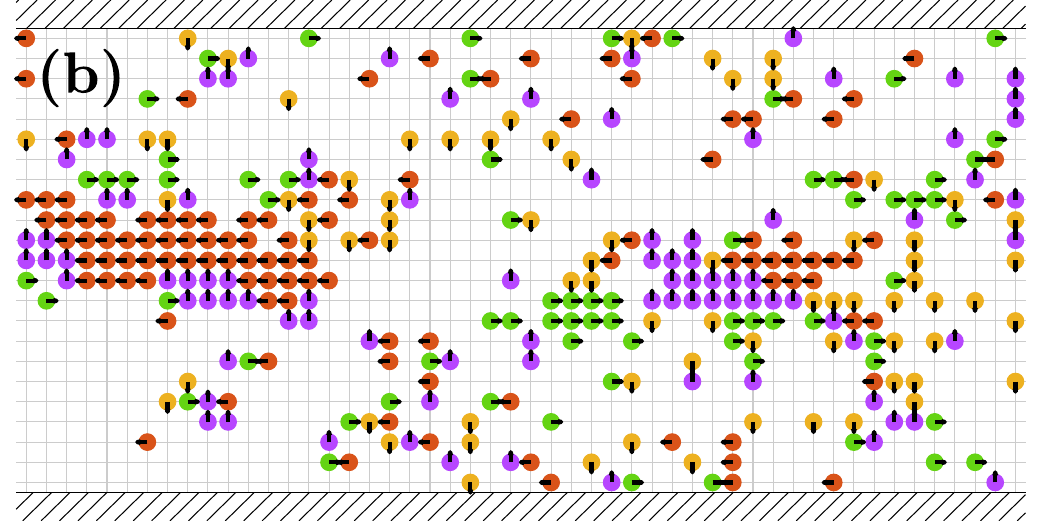}\\
	\includegraphics[width=.49\columnwidth]{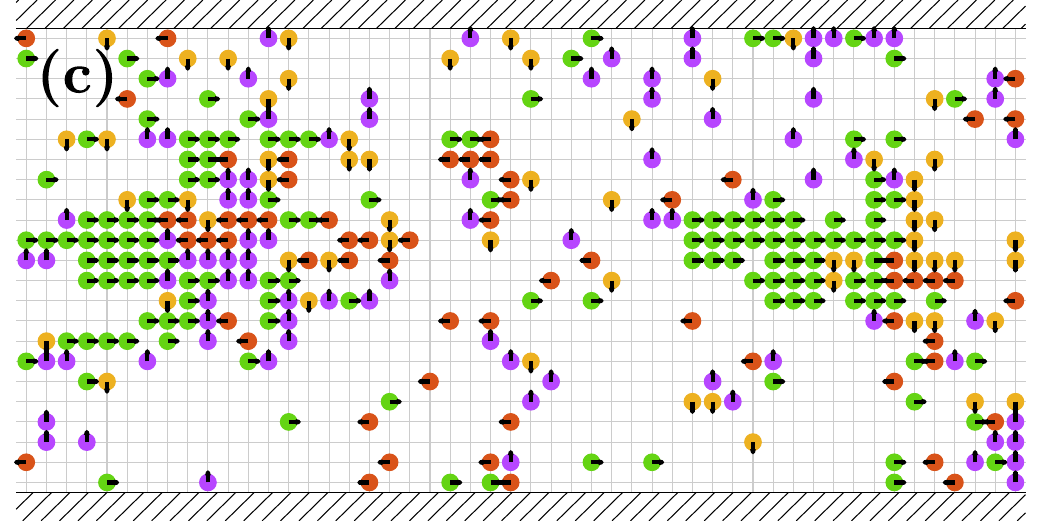}
	\includegraphics[width=.49\columnwidth]{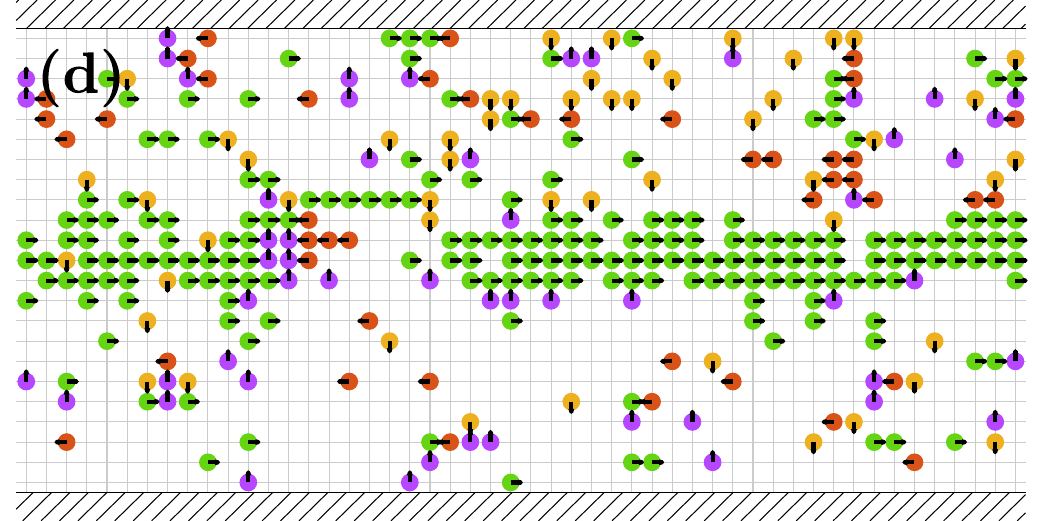}
	\caption{\label{fig:snaps_phases_v1_40} \textit{Typical lattice configurations in the intermediate phase}, shown for $U/v_{\rm S}=0.4$ and \textbf{(a)}~$g=1.4$, \textbf{(b)} $g=1.5$, \textbf{(c)} $g=1.6$, and \textbf{(d)} $g=1.7$, all within the interval $[g_{\rm o}^\star,g_{\rm b}^\star]$.}
\end{figure}
 \begin{figure}[b]
	\centering
	\includegraphics[width=\columnwidth]{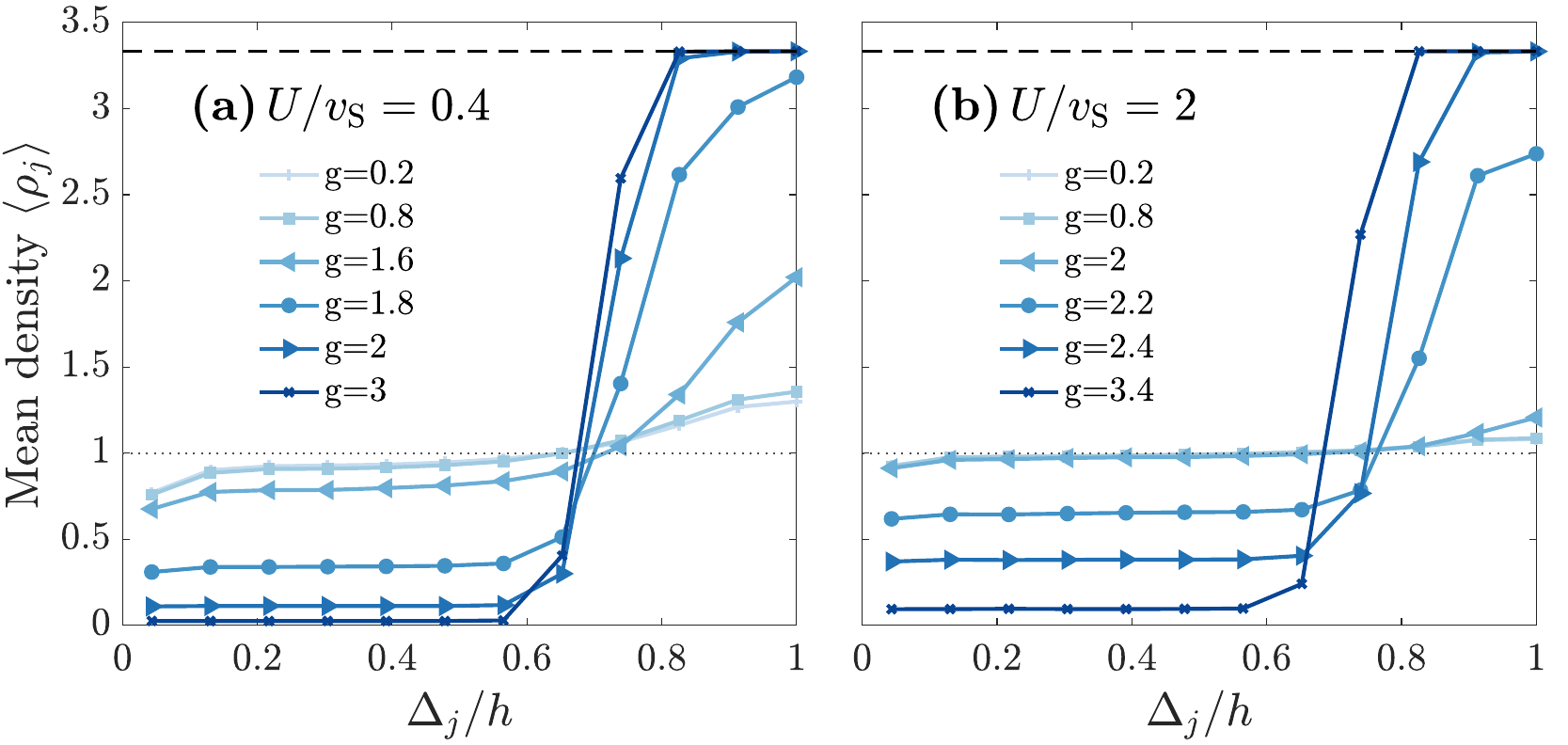}
	\caption{\label{fig:density_flow} \textit{Density profiles as a function of the distance $\Delta_j$ from the boundary} for two different fluid velocity amplitudes. The alignment sensitivity $g$ spans values both below and above the transition. The dotted line represents the uniform distribution and the dashed line dense packing.}
\end{figure}
In this intermediate phase, the particles self-organize into patterns that differ significantly from the clogs and obstructions observed in the absence of fluid flow. Figure~\ref{fig:snaps_phases_v1_40} shows instantaneous lattice configurations for four different values of the alignment sensitivity $g$ lying in between $g_{\rm o}^\star$ and $g_{\rm b}^\star$. In all cases, particles form \textit{medial clumps} predominantly located in the center of the channel, rather than adhering to the walls. We attribute this tendency to the fact that the flow vorticity is minimal and vanishes along the central axis of the channel, while shear is most pronounced near the boundaries. Consequently, particles close to the walls experience stronger angular dynamics, which disrupts their ability to maintain orientational order and to form structures that are attached to the walls. This tendency is more evident from Fig.~\ref{fig:density_flow}, which shows average density profiles as a function of the distance from the boundary. The increased concentration of particles in the channel's core is actually evident across all alignment sensitivities: it is present even for $g<g_{\rm o}^\star$ in the absence of order, and persists for $g>g_{\rm b}^\star$, when particles organize into streamwise bands.

\begin{figure}[h]
	\centering
	\includegraphics[width=.49\columnwidth]{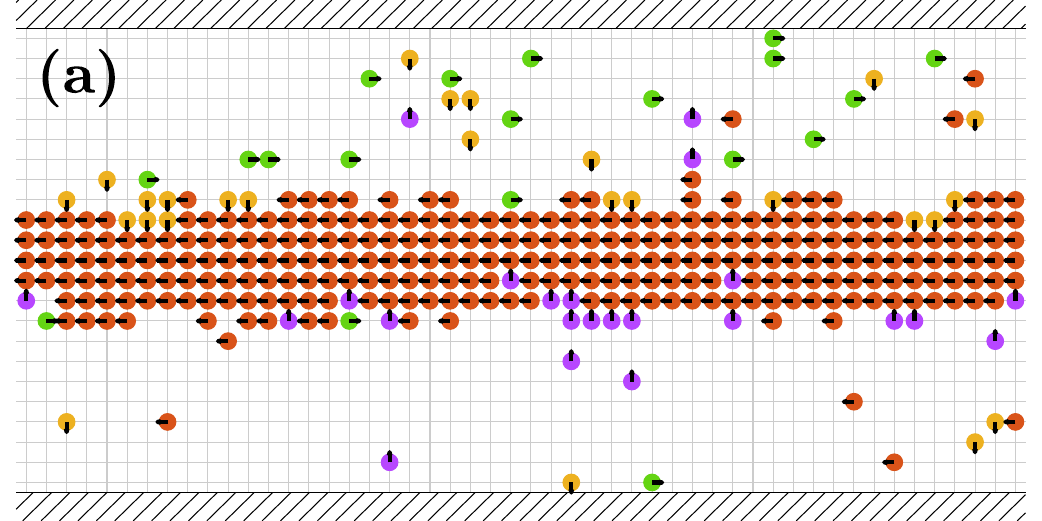}
	\includegraphics[width=.49\columnwidth]{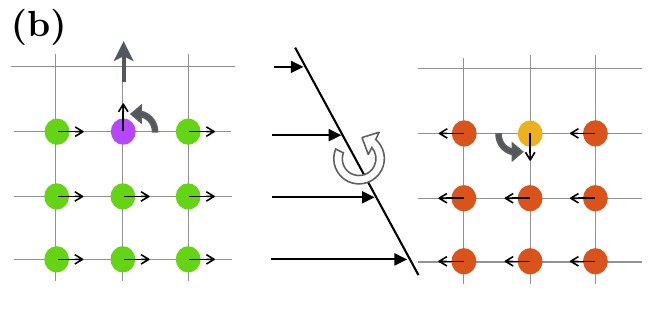}
	\caption{\label{fig:back_facing} \textbf{(a)}~Typical lattice configuration in the ordered band phase obtained for $U/v_{\rm S}=0.4$ and $g=1.8$. \textbf{(b)}~Sketch of the mechanism by which a downstream-oriented band (left) is less stable than an upstream-oriented one (right).}
\end{figure}
The tendency of particles to concentrate in the region where the fluid flow is the strongest might a priori seem to favor their transport, particularly when they align with the flow direction, as seen for instance in Fig.~\ref{fig:snaps_phases_v1_40}d. However, our simulations reveal that in the ordered phase, particles consistently orient themselves facing upstream, effectively counteracting by their active mobility the advection by the flow. An illustration is given in Fig.~\ref{fig:back_facing}a, which shows a typical configuration obtained for $g>g_{\rm b}^\star$. Qualitatively, this preferential orientation arises due to the particle tumbling induced by the fluid vorticity. As described in Sec.~\ref{sec:model}, the flow rotates particles either clockwise or counterclockwise, depending on which half of the channel they are located in. Without loss of generality, let us consider the situation in the upper half of the channel ($y>0$), as sketched in Fig.~\ref{fig:back_facing}b. For a downstream-oriented band (comprising \rightcirc\ particles aligned toward $x>0$), the vorticity causes the particles at the edge of the band to rotate in the positive direction, allowing them to potentially escape it. In contrast, in an upstream-oriented band (comprising  \leftcirc\ particles aligned toward $x<0$), the rotation induced by the flow does not have such a disruptive effect; the rotated particle is now constrained by its neighbors, maintaining the band's cohesion. Consequently, downstream-oriented bands are more likely to losing particles and are therefore less stable compared to upstream-oriented patterns.

\begin{figure}[h]
   	\centering
   	\includegraphics[width=\columnwidth]{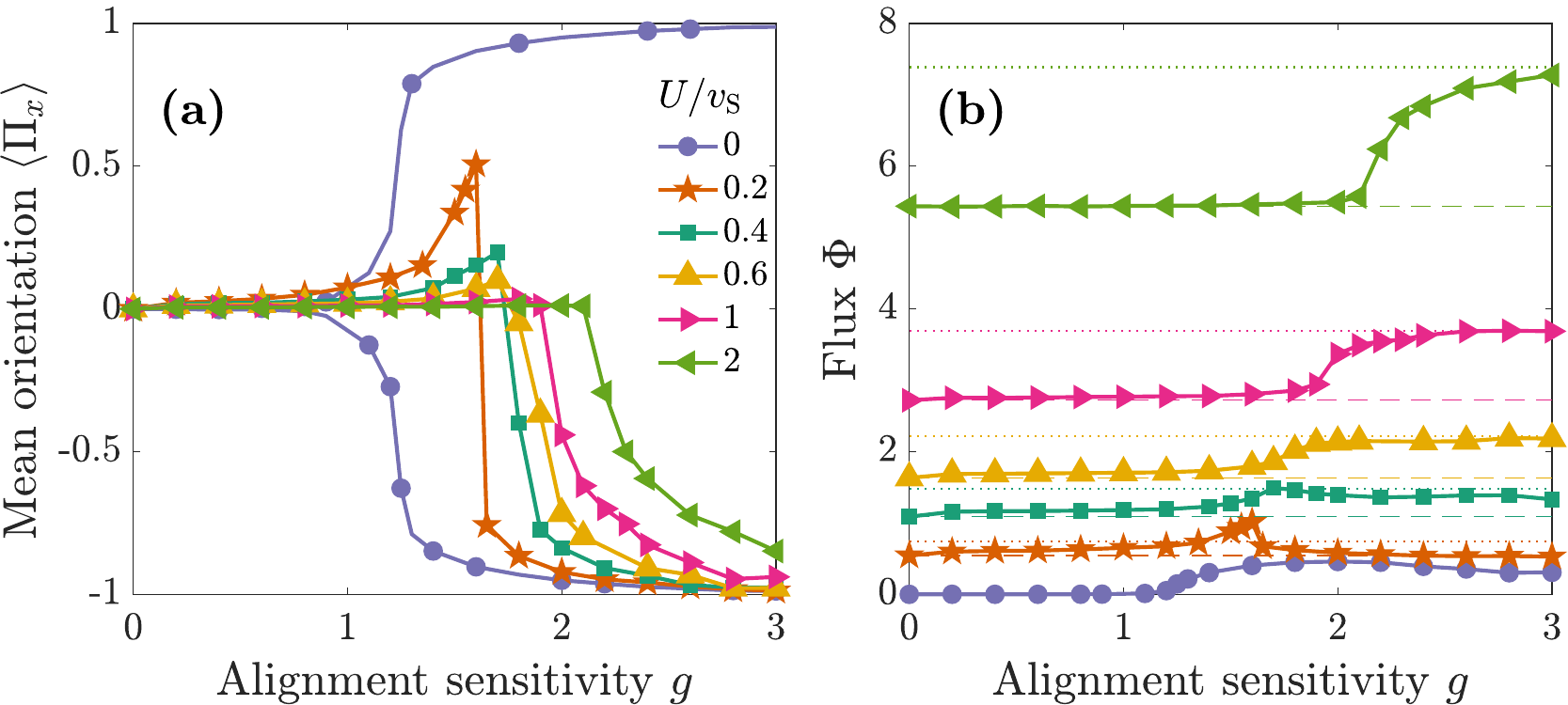}
	\caption{\label{fig:alignment_flux_with_flow} \textbf{(a)} Mean orientation in the streamwise direction as a function of $g$ for various fluid-flow velocities. In the case without fluid flow ($U=0$), the solid lines represent $\pm\langle|\Pi_x|\rangle$, while the symbols indicate the actual realization-dependent sign of $\Pi_x$, observed at the end of our simulations. \textbf{(b)} Corresponding flux values. For $U=0$, absolute values have been used. The horizontal dashed lines indicate the flux induced by the fluid flow on homogeneously distributed particles with random orientations. The horizontal dotted lines are the asymptotic flux associated to a densely packed, centered band (see text).}
\end{figure}
The preferential alignment of particles in the presence of a Poiseuille flow can be assessed by measuring the average streamwise component $\langle\Pi_x\rangle$ of the mean orientation vector. The dependence of this quantity upon $g$ and $U/u_{\rm S}$ is shown in Fig.~\ref{fig:alignment_flux_with_flow}a.  When $U=0$, the system is symmetric under $x\mapsto-x$, so that $\langle\Pi_x\rangle=0$. Hence, we rather plot in this case $\pm\langle|\Pi_x|\rangle$ to facilitate comparison. When $U>0$, the fore-aft symmetry is broken, giving $\langle\Pi_x\rangle$ a definite sign. Measurements clearly indicate that at all values of $U/v_{\rm S}$ and for $g$ values where an ordered band phase is achieved, the particles systematically form upstream-facing patterns with $\langle\Pi_x\rangle<0$. Interestingly, the opposite behavior is observed for alignment sensitivities below $g_{\rm b}^\star$. In this regime, particles tend to align in the same direction as the fluid velocity, making the observation in Fig.~\ref{fig:snaps_phases_v1_40}d a representative instance of the actual patterns developed in the system.

Our interpretation of this downstream preferential alignment is that, at intermediate values of $g$, the structures in which particles self-organize are not actual bands, but rather involve continuous exchanges with particles that travel back and forth from the boundaries. These medial clumps are hence constantly fed with \upcirc\ particles in the lower half of the channel and \downcirc\ particles in its upper half. Once collected, these particles are stacked until they reorient by half a turn and leave the aggregate. When this reorientation process is primarily driven by the flow, particles tumble clockwise in the lower half of the channel and counterclockwise in the upper half, inevitably leading to a stage where they align as \rightcirc. This preferential orientation thus arises from a dynamical bias, in contrast to the stability-driven alignment observed in band structures.

We next examine how preferential concentration in the channel center, pattern formation, and alignment with or against the fluid flow combine or compete to influence particle transport properties. Figure~\ref{fig:alignment_flux_with_flow}b shows the average flux $\Phi$ as a function of $g$ for various values of $U/v_{\rm S}$. To again facilitate comparisons, we also plot the mean flux amplitude $\langle|\Phi|\rangle$ in the case without a fluid flow. At low alignment sensitivities, the flux essentially matches that of randomly oriented particles uniformly distributed across the channel (as indicated by horizontal dashed lines). In the intermediate regime $g_{\rm o}^\star<g<g_{\rm b}^\star$, the combination of concentration in the center of the channel and preferential downstream alignment leads to an increase in the mean flux, particularly at low to moderate values of $U/v_{\rm S}$. At higher $g$ values within the orientationally ordered phase, there is a balance between particle concentration in high-velocity regions and upstream particle motion due to backward alignment. However, as already observed in the absence of fluid flow, particles within bands are densely packed and single-occupancy constraints on the lattice drastically inhibits their active mobility. Consequently, preferential concentration becomes dominant, as evidenced by the observed increase in flux with increasing $g$. At very large sensitivities, the flux reaches an asymptotic value corresponding to a fully packed band, centered in the channel and with a width $\approx 2h\rho$ determined by the global density. The asymptotic flux is then given by integrating the fluid velocity on $-h\rho< y<h\rho$, leading to $\Phi \propto U\,\rho(1-\rho^2)/3$ (shown as dotted lines on the plot). In conclusion, these observations highlight that particle transport properties are significantly influenced not only by self-orientation but also by their spatial distribution within the domain.


\section{Concluding remarks}
\label{sec:conclusions}

In this study, the self-organization of self-propelled particles subjected to a Poiseuille flow within a confined channel was numerically investigated using a discrete Vicsek-like interaction model on a single-occupancy square lattice. This model allowed for a detailed exploration of the phases that emerge\,---\,from disordered to orientationally ordered regimes\,---\,and their impact on particle transport, particularly focusing on the flux of particles moving in the flow direction.

The examination of the system without fluid flow revealed that while a confined geometry does not significantly alter the phase transition compared to unbounded settings, it introduces wall-induced particle accumulations. This results in the formation of clogs and bands aligned with the walls, which can impede particle movement. Despite these dense clusters, the organization of particles into band structures along the channel still allows for an overall flux. It was also found that larger domain aspect ratios introduce new spatial structures, such as obstructions extending between the two channel walls, and lead to longer transient states and the need for higher alignment sensitivities to achieve order. We noted that such structures can exhibit prolonged stability before transitioning to a steady state and contribute to highly intermittent fluctuations in the overall flux.

In the presence of Poiseuille flow, particles experience flow-induced vorticity that interacts with their alignment tendencies, weakening orientational interactions and thereby shifting the phase transition to higher values of the alignment sensitivity, akin to adding orientational disorder to the system. Interestingly, particles tend to form clusters away from the walls, concentrating at the channel center, even for alignment sensitivities below the transition to band patterns. This preferential concentration in regions of maximal fluid velocity is responsible for a significant improvement in the overall particle flux, despite the fact that particles forming bands align against the flow, counteracting their transport.

\

A key aspect of our study is the use of a square lattice with edges aligned with the channel walls. This discretization introduces anisotropies that could influence the observed collective pattern formation and stability, as well as their implications for particle transport. Fully understanding the impact of this discretization requires exploring alternative lattice structures, such as triangular or hexagonal lattices, or employing semi-discrete models as in~\cite{kuhn2021lattice}. Additionally, our simulations use reflective boundary conditions, which tend to impart a repulsive nature to the walls. Exploring alternative boundary conditions that more closely reflect phenomena such as wall trapping observed for micro-swimmers~\cite{bechinger2016active} could provide more realistic insights into the particle self-organization.

Furthermore, it is important to notice that deriving continuous hydrodynamic limits for the system studied in this paper could provide valuable analytical insights. Previous studies on simpler systems, such as the active Ising model with no exclusion~\cite{chatterjee2020flocking} or exclusion processes with slightly simpler orientational interactions~\cite{kourbane2018exact}, have demonstrated the utility of hydrodynamic approaches for understanding collective dynamics. Applying similar techniques to this more complex system could enhance our theoretical understanding and provide predictions regarding transitions and the characterization of the different regimes.

Let us finally mention that overall, our findings underscore the potential to leverage self-organization properties to enhance transport efficiency in confined channels. These insights suggest new avenues for controlling swarms of micro-swimmers in practical applications, such as targeted drug delivery or environmental monitoring. By understanding and manipulating the self-organization dynamics and associated transport properties, it may be possible to design systems with optimized performance for specific displacement tasks. This study is hence contributing to our long-term research agenda, which focuses on leveraging self-organization and spontaneous pattern formation to optimize the navigation of swarms of microswimmers.

\begin{acknowledgments}
	Computational resources were provided by the OPAL infrastructure from Université Côte d’Azur. We received support from the UCA-JEDI Future Investments, funded by the French government (Grant No.~ANR-15-IDEX-01), from the Agence Nationale de la Recherche (Grants No. ANR-21-CE45-0013 \& ANR-21-CE30-0040-01), and from the European Research Council (ERC) under the European Union’s Horizon 2020 research and innovation programme (Grant Agreement No. 882340).
	
	We acknowledge financial support under the National Recovery and Resilience Plan (NRRP), Mission 4 Component 2 Investment 1.1 -- Call for tender No.\ 104 published on date 02/02/2022 by the Italian Ministry of University and Research (MUR), funded by the European Union---NextGenerationEU---Project Title: Equations informed and data-driven approaches for collective optimal search in complex flows (CO-SEARCH), Concession Decree No. 957 of 30/06/2023, Project code 202249Z89M--CUP E53D23001610006.
\end{acknowledgments}

\end{document}